**Machine learning of neuroimaging to diagnose cognitive impairment and dementia: a systematic review and comparative analysis.**

**Enrico Pellegrini PhD[1], Lucia Ballerini PhD[1], Maria del C. Valdes Hernandez PhD[1,] Francesca M. Chappell PhD[1], Victor González-Castro PhD[2], Devasuda Anblagan PhD[1], Samuel Danso PhD[1], Susana Muñoz Maniega PhD[1], Dominic Job PhD[1], Cyril Pernet PhD[1], Grant Mair MB ChB, FRCR, MD,[1], Tom MacGillivray PhD[1,3], Emanuele Trucco PhD[4], Joanna Wardlaw MB ChB, FRCR, MD[1,5]**

[1]Division of Neuroimaging, Centre for Clinical Brain Sciences and Edinburgh Imaging, University of Edinburgh, UK

[2]Department of Electrical, Systems and Automatics Engineering, Universidad de León, León (Spain)

[3] VAMPIRE project, University of Edinburgh, UK

[4]VAMPIRE project, Computing, School of Science and Engineering, University of Dundee, UK

[5] UK Dementia Institute at the University of Edinburgh, UK

**Correspondence**: Joanna Wardlaw, joanna.wardlaw@ed.ac.uk

**ABSTRACT**

INTRODUCTION: Advanced machine learning methods might help to identify dementia risk from neuroimaging, but their accuracy to date is unclear.

METHODS: We systematically reviewed the literature, 2006 to late 2016, for machine learning studies differentiating healthy ageing through to dementia of various types, assessing study quality, and comparing accuracy at different disease boundaries.

RESULTS: Of 111 relevant studies, most assessed Alzheimer's disease (AD) vs healthy controls, used ADNI data, support vector machines and only T1-weighted sequences. Accuracy was highest for differentiating AD from healthy controls, and poor for differentiating healthy controls vs MCI vs AD, or MCI converters vs non-converters. Accuracy increased using combined data types, but not by data source, sample size or machine learning method.

DISCUSSION: Machine learning does not differentiate clinically-relevant disease categories yet. More diverse datasets, combinations of different types of data, and close clinical integration of machine learning would help to advance the field.

## INTRODUCTION

Ageing is associated with increasing health care costs of which two related neurological disorders, dementia and stroke, account for much of the increase. Dementia is a progressive development of multiple cognitive deficits with several underlying aetiologies, the two commonest types being Alzheimer's disease (AD) and vascular dementia (VaD). The total estimated worldwide cost of dementia was US$818 billion in 2015, representing 1.09% of global GDP.[1] In 2015, 46.8 million people worldwide were living with dementia, a figure which is expected to almost double every 20 years, reaching 74.7 million in 2030 and 131.5 million by 2050. Meanwhile, stroke remains the second commonest cause of death and commonest cause of dependency in adults worldwide.[2]

Age-related cognitive decline ranges from minor reductions in memory and executive function that do not interfere with daily life, to more severe degrees that fall short of dementia but may interfere with some activities of daily living, termed 'mild cognitive impairment'. Mild cognitive impairment (MCI) may progress to dementia or remain static, and cognitive decline is also a risk factor for stroke.

All three of MCI, dementia and stroke are associated with changes seen on brain imaging particularly brain volume loss (atrophy) and development of focal lesions in the white and grey matter such as white matter hyperintensities (WMH), lacunes, microbleeds, focal cortical or subcortical infarcts or small haemorrhages. These features are also associated with ageing (though are less frequent in healthy ageing), may be symptomatic or asymptomatic, and predict increased risk of stroke, dementia and death.[3]

In the last decade, improvements in medical imaging, higher image quality, the exponential increase in computational power of affordable computing platforms, and the greater availability of brain imaging datasets such as from the Alzheimer's Disease Neuroimaging Initiative (ADNI), have increased opportunities to develop machine learning approaches aimed at the automated detection, classification and quantification of diseases.[4] Some of these techniques have been applied to classify brain magnetic resonance imaging (MRI) or computed tomography (CT) scans, comparing patients with dementia and healthy controls and to distinguish different types or stages of dementia, cerebrovascular disease and accelerated features of aging. However, the recent rapid increase in publications using different machine learning techniques in different populations, types of images and disease criteria, make it difficult to obtain an objective view of the current accuracy of machine learning.

We undertook this systematic review to critically appraise the accuracy of machine learning to differentiate healthy ageing from mild cognitive impairment from dementia and predict the future risk of dementia or cerebrovascular disease. We evaluated the performance metrics of individual machine learning techniques by task, disease of interest, imaging sequence and features investigated.

## METHODS

### Search Strategy

We searched the literature from 1st Jan 2006 (when first publications on machine learning in the disorders of interest started appearing in earnest) to 30th September 2016, on six databases: Pubmed/Medline, Elsevier, IEEE Xplore Digital Library, Science Direct, ACM Digital Library and Web of Science.

We devised three groups of keywords, each relevant to different aspects of the scope of the review:

*Brain lesions and relevant pathologies*: Dement*, Alzheimer, AD, VCI, VaD, small vessel disease, SVD, microvascular change, cognitive impairment, cognitive decline, MCI, Lewy bod*, LBD, frontotemporal, FTD, lacun*, white matter hyperintens*, white matter lesion*, WMH, leukoaraiosis, periventricular, microbleed*, microhaemorr*, microhemorr*, stroke, cerebrovascular, CVA, perivascular space*, PVS, Virchow–Robin space*, pathological aging, pathological ageing, brain, cerebr*, medial temporal, mesial temporal, volume loss, atrophy.

*Machine learning:* machine learning, supervised learning, unsupervised learning, deep learning, classification, identification, detection, automat* diagnosis, pattern analysis, CAD, computer aided diagnosis, computer assisted diagnosis, computational analysis.

*Structural imaging:* MR, magnetic resonance, structural imag*, CT, CAT, computed tomograph*.

We searched titles, abstracts and keyword fields of indexed studies, published as journal papers or conference proceedings, with all possible strings obtained by joining one term from each of the above groups with an "AND" operator. One reviewer (EP) conducted the searches and eliminated all duplicate references.

**Inclusion/Exclusion Criteria**

Two reviewers (EP, VGC) separately assessed all non-duplicate papers in a two-stage selection process. First, we evaluated titles and abstracts to exclude studies clearly not relevant to the scope of the review. Second, we assessed full texts of the remaining papers to eliminate studies using the following exclusion criteria:

1. Studies of animals or ex-vivo samples

2. Reviews, surveys, collections and comparison papers not presenting a new ML method or application.

3. Studies with a validation set comprising a small number of subjects (<100 for disease classification or lesion identification tasks, and <25 for pixel or voxel level lesion segmentation tasks) or with a manual ground truth provided by only one trained observer.

4. Studies presenting a method in which the main task (e.g., lesion segmentation) was not performed in a fully automated fashion. Studies involving semi-automated pre-processing steps (e.g., brain parcellation refinement) obtained by making use of previously validated software and trained observers were accepted.

5. Studies not about structural MRI or CT imaging.

6. Studies focused on image pre-processing techniques that did not include any machine learning for disease classification or lesion segmentation/identification (e.g., contrast enhancement, noise reduction techniques, etc.).

7. Studies of parcellation of healthy brain regions not used for disease classification or detection.

8. Studies that either did not provide, or presented their results in such a way that we were not able to calculate performance metrics (e.g. sensitivity and specificity).

9. Multiple publications from the same research group, focusing on the same task and dataset. In such cases, only the most recent publication or with the largest sample size was included in the data analysis.

10. Studies that did not describe their methods in sufficient detail to enable replication.

Discrepancies were resolved by discussion between the two reviewers with a third (MvH, LB, GM) arbitrating as necessary.

**Data Extraction**

From the included papers, we extracted data on the:

(1) disease or lesion investigated,

(2) dataset used and whether it was publicly available or not,

(3) number of subjects or images on which the proposed technique had been validated,

(4) type of structural imaging modality  and sequences used,

(5) imaging features that were investigated,

(6) use of any additional imaging data (e.g., functional imaging) or non-imaging features (e.g., cognitive test scores) in the analysis,

(7) classifier(s) and the feature selection and representation techniques used, and

(8) performance (sensitivity, specificity, accuracy) of the proposed method.

We extracted data to calculate sensitivity and specificity where not already calculated.

If multiple tasks were investigated in a single study, the respective data for each experiment were recorded.

We also extracted (when reported) details of: use of single vs multiple scanners, image resolution, population demographics, exclusion criteria for each dataset, image pre-processing steps, time cost, and use of third party software (details available on request).

We evaluated study quality according to the relevant Quality Assessment of Diagnostic Accuracy Studies 2 (QUADAS-2) criteria ([https://www.ncbi.nlm.nih.gov/pubmed/22007046). We used the seven criteria that were most relevant to the material of the review, four addressing risk of bias and three addressing applicability. since some criteria were not strictly applicable to the field.

All acronyms used in the results table are reported in Supplemental Table 1.

**Data Analysis**

We extracted the different performance metrics directly from the papers, or calculated them from the data provided. In particular, we aimed to examine:

1. Sensitivity, specificity and accuracy for binary classification tasks.

2. Mean class accuracy for multi-class classification tasks.

3. Dice coefficient (DC) for accuracy of lesion segmentation tasks.

4. Precision and recall for lesion identification tasks (calculated using the formula in Supplementary methods).

Where the results of multiple experiments for the same classification task were reported in a single study, we only used the set of metrics associated with the higher value of accuracy in our analysis.

We constructed forest plots to summarise sensitivity, specificity, accuracy and 95% confidence intervals

(CI) of various clinically relevant diagnoses including AD versus healthy ageing, MCI versus AD or healthy ageing, MCI conversion to AD versus not conversion. In order to summarise the mass of information effectively, we plotted forest plots of accuracy rather than sensitivity and specificity, defined as:

$$Accuracy = \frac{TP + TN}{TP + FN + TN + FP}$$

We performed sensitivity analyses to determine if source dataset, machine learning method, type of data used, or study size accounted for the variance between studies. We calculated 95% CI of accuracy using the Wilson score method. We plotted all graphs in R. We considered but rejected performing a formal meta-analysis, since the huge overlap in datasets in publications precluded determining the results of patients who contributed to more than one study (even with exclusion of obvious duplicate publications), preventing the modelling of between-study variance. Finally, to minimise confounding by inclusion of studies that only contributed to one comparison, we compared accuracy across multiple diagnostic boundaries using studies that provided data on more than one diagnostic comparison from the same dataset.

**Role of the Funding Source**

The funders had no role in the conduct of this systematic review. The corresponding author confirms that she had full access to all the data in the study and had final responsibility for the decision to submit for publication.

**RESULTS**

Our search yielded 5775 non-duplicate studies, of which 4978 (86%) were excluded at title/abstract screening as clearly not relevant to the review. After full text screening, we found 111 papers relevant for data extraction (Figure 1). The two criteria accounting for the most exclusions were small sample (item 3) and no performance metrics provided or calculable (item 8; respectively 41% and 19% of exclusions at this stage; proportions meeting exclusion criteria see Supplementary Table 2). Note that studies that failed one exclusion criterion were excluded and not evaluated further; although some might have failed on multiple criteria, we only recorded the first reason for exclusion.

Most of the 111 studies that met inclusion criteria achieved low risk of bias scores and low concerns on applicability (Supplement Figure 1). Of the 111, we used 89 studies in further analyses of accuracy where data could be extracted as 2x2 tables and there were sufficient studies to compare.

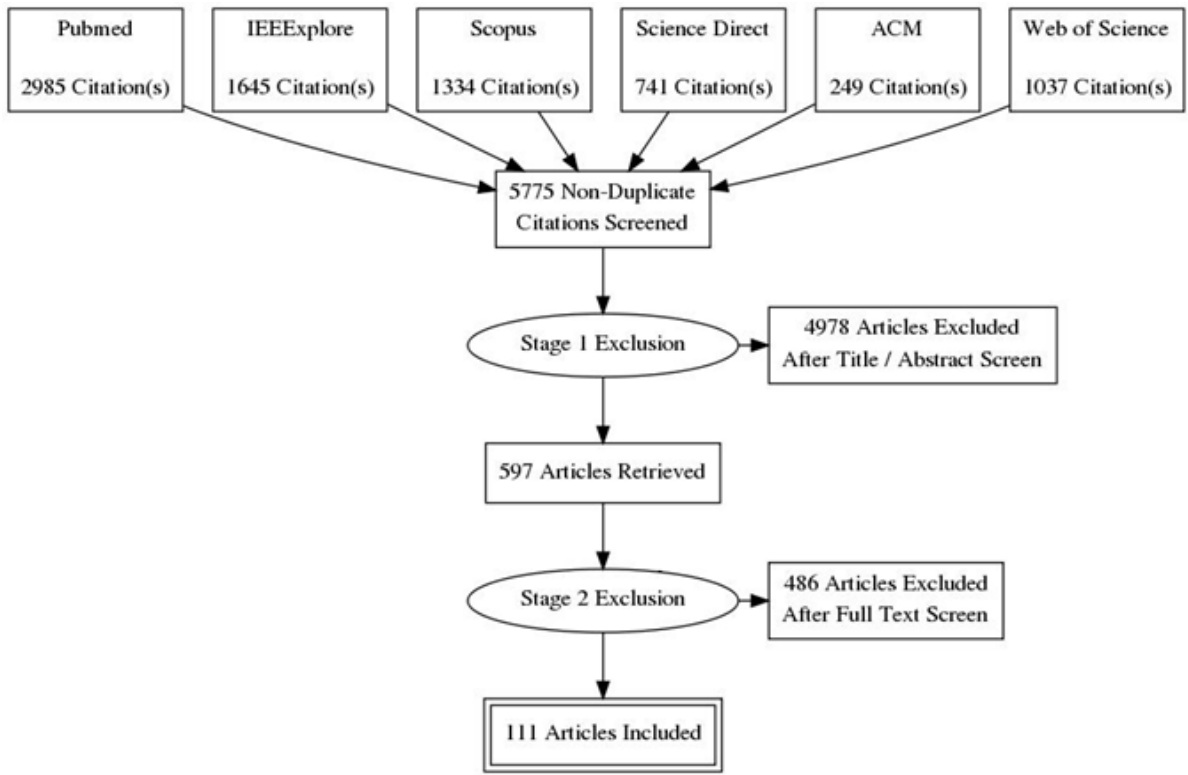

**Figure 1** Flowchart of search and exclusion stages of the review.

Most studies tested the diagnosis of AD (68/89, 76%), most versus healthy controls (67/89, 75%), then MCI non-converters to AD versus converters to AD (37/89, 42%), MCI versus healthy controls (29/89, 33%), and MCI versus AD (8/89, 9%; Table 1 shows individual comparisons; full details in Supplementary Table S3). There were 21 studies that compared multiple diagnostic classes, of which many involved the same author groups.

The remaining studies focused on other factors: other types of dementia (five studies, Supplementary Table S4), and studies investigating different types of brain lesions related to dementia, stroke and pathological aging, either: lesion segmentation (seven studies, Supplementary Table S5) or lesion identification (11 studies, Supplementary Table S6). As there were few eligible studies in the latter three categories, it was not possible to undertake any formal comparisons, e.g. of DICE coefficients (for WMH, ischaemic stroke lesions), Precision or Recall values (for microbleeds, lacunes). However the DICE coefficients for WMH segmentation (four studies, mean n=81, range 38-125) ranged from 0.520-0.691 and for infarcts (three studies, mean n=42, range 30-60) ranged from 0.670-0.740 (Supplementary Table S5). The Precision/Recall values for microbleeds (three studies, mean n=66, range 50-81) for Precision were 0.101-0.443 and for Recall were 0.870-0.986; there was one study on lacunes (n=132) with Precision of 0.154 and Recall of 0.968 (Supplementary Table S6).

| Data sources | HC v AD | HC v MCI | MCInc v MCIc | MCI v AD | Total |
|---|---|---|---|---|---|
| ADNI | 54 | 24 | 34 | 7 | 119 |
| ADNI + Bdx-3C | 0 | 0 | 1 | 0 | 1 |
| AddNeuroMed | 1 | 0 | 2 | 0 | 3 |
| AddNeuroMed + ADNI | 2 | 1 | 1 | 0 | 4 |
| Local | 4 | 3 | 0 | 0 | 7 |
| OASIS | 7 | 2 | 0 | 1 | 10 |
| Total | 68* | 30 | 38 | 8 | 144 |
| **Machine learning method** | | | | | |
| AdaBoost | 1 | 0 | 1 | 0 | 2 |
| Deep Learning | 2 | 2 | 0 | 0 | 4 |
| Gaussian Process | 0 | 0 | 1 | 0 | 1 |
| LDA | 5 | 0 | 5 | 1 | 11 |
| Logistic Regression | 4 | 0 | 2 | 0 | 6 |
| OPLS | 2 | 1 | 1 | 0 | 4 |
| QDA | 0 | 0 | 1 | 0 | 1 |
| RBF-NN | 0 | 0 | 1 | 0 | 1 |
| Random Forest | 3 | 1 | 3 | 0 | 7 |
| SRC | 2 | 1 | 2 | 0 | 5 |
| SVM | 39 | 22 | 17 | 7 | 85 |
| SVM + MKL | 3 | 1 | 1 | 0 | 5 |
| SVM + OPLS | 1 | 0 | 1 | 0 | 2 |
| SVM + Random Forest | 2 | 1 | 2 | 0 | 5 |
| SVM + SRC | 1 | 1 | 0 | 0 | 2 |
| kNN | 3 | 0 | 0 | 0 | 3 |
| Total | 68* | 30 | 38 | 8 | 144 |
| **Types of Imaging and imaging plus non-imaging data used** | | | | | |
| T1w only | 46 | 13 | 26 | 6 | 91 |
| T1w & other imaging data | 8 | 8 | 2 | 0 | 18 |
| T1w & other types of data | 8 | 3 | 8 | 1 | 20 |
| T1w & both other imaging & types of data | 6 | 6 | 2 | 1 | 15 |
| Total | 68* | 30 | 38 | 8 | 144 |
| **Size of dataset (range from 100 to 902 participants).** | | | | | |
| 150 and under | 30 | 4 | 9 | 2 | 45 |
| 151 to 200 | 4 | 10 | 6 | 0 | 20 |
| 201 to 250 | 9 | 4 | 6 | 0 | 19 |
| 251 to 300 | 4 | 2 | 3 | 0 | 9 |
| Over 300 | 21 | 10 | 14 | 6 | 51 |
| Total | 68* | 30 | 38 | 8 | 144 |

**Table 1** Number of comparisons in each systematic review analysis group using specified data source, machine learning method, types of imaging and non-imaging data and by study size. Individual studies contribute to more than one analysis and use more than one data source, machine learning method, combinations of imaging data and more than one dataset (hence more than one sample size in some studies). HC=healthy control; AD=Alzheimer's disease; MCI=mild cognitive impairment; nc=non converter to AD; c=converter to AD.

The 76 analyses focused on AD (Supplementary Table S3) amounted to 68 unique references, with huge overlap in authors and data sources between the studies. As well as using more than one data source, many studies performed more than one comparison of disease classifications with these multiple data sources, hence amounting to 144 different comparisons (Table 1). Of the 144 comparisons, there were 120 uses of ADNI data (ADNI alone 119/144, 83%; ADNI plus other 120/144, 83%), followed by Oasis (10/144, 7%), local sources (7/144, 5%), and AddNeuroMed (alone 3/144, 2%; plus ADNI 4/144, 3%).

The 76 analyses of AD tested nine different machine learning methods. The most frequent, by a large margin, was Support Vector Machine (SVM) with 46/76 (61%) alone and 53/76(70%) combined with another machine learning method, then linear discriminant analysis (LDA, 6/76, 8%), logistic regression (4/76, 5%) and a few testing k-nearest neighbours (KNN), Orthogonal Projections to Latent Structures (OPLS), Random forest, or Sparse Representation Classification (SRC), Table 1. Most analyses, by a large margin, used only T1 images (91/144, 63%), with modest numbers using T1 plus other sequences, other types of data, or both. Analysis sample sizes ranged from 100 to 902, with similar numbers of analyses including more than 300 subjects (51/144, 35%) or fewer than 150 subjects (45/144, 31%), Table 1.

Amongst the 76 studies focused on AD, the accuracy was higher for differentiating AD from healthy controls (most study accuracies were in the 0.8-1.0 range), than for differentiating MCI from healthy controls (accuracies =0.6-0.9), or non-converting from converting MCI to AD (accuracies= 0.5-0.85), or MCI from AD (accuracies =0.6-0.9). Figure 2a-d indicates the lower accuracy for differentiating healthy controls from MCI, or MCI from AD, or MCI non-converters from converters, than healthy controls from AD; Supplementary Figures 2-4 illustrate these same comparisons ordered by data source, machine learning method and study size respectively. There was little evidence of any difference in accuracy by machine learning method, data source used, or study size, with possible higher accuracy for combined T1 plus other sequences and other types of data than for T1 imaging alone.

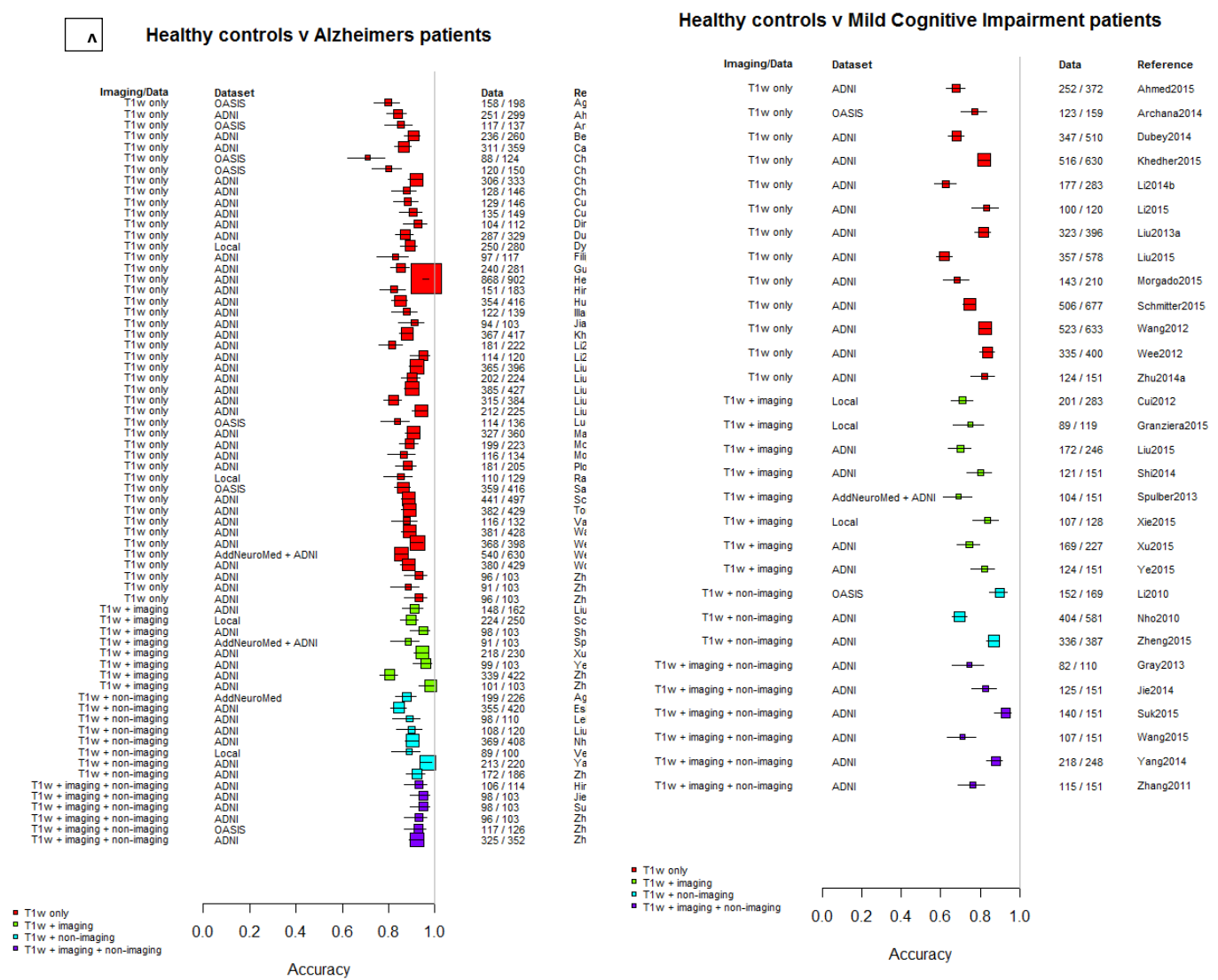

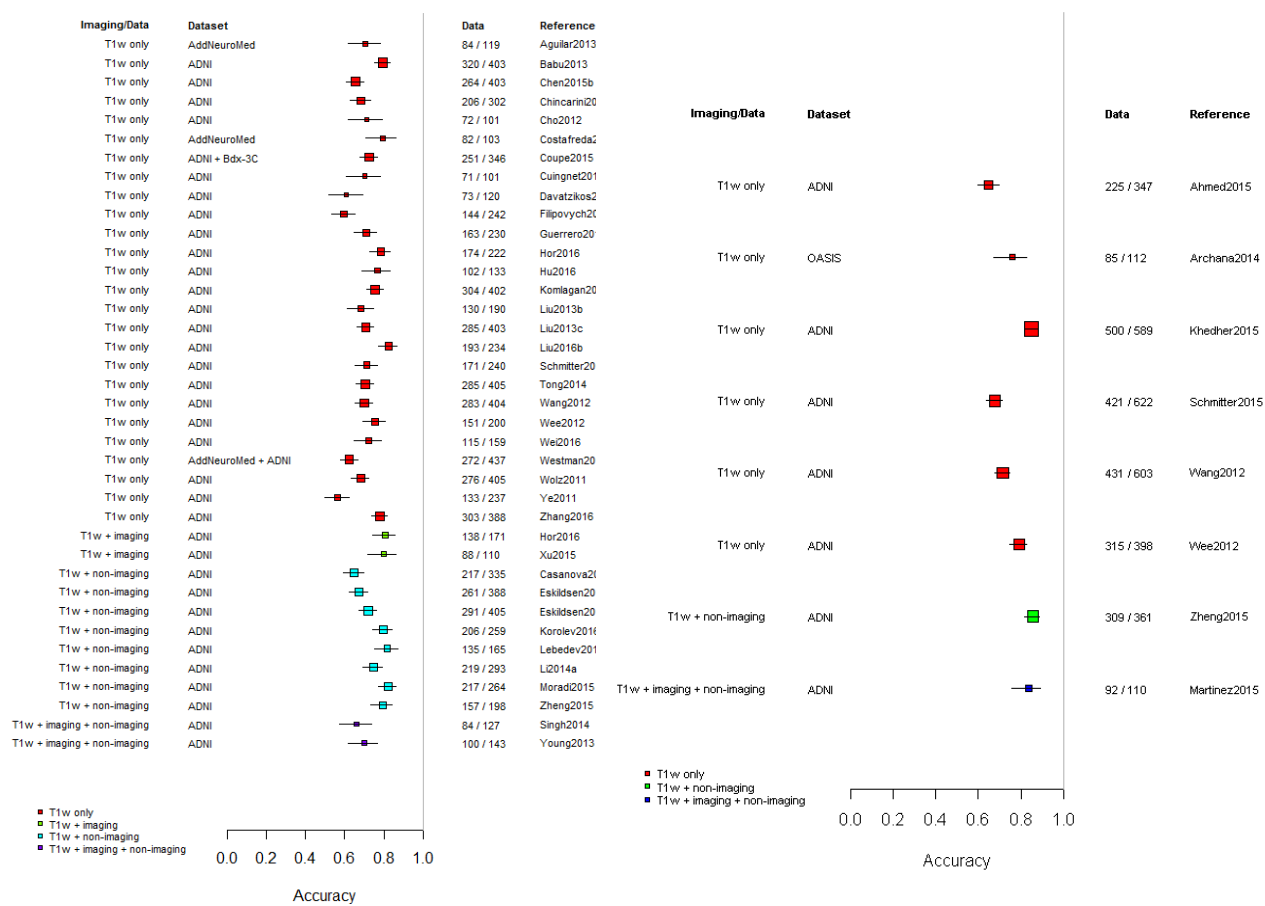

**Figure 2.** Differentiation of a) healthy controls from AD, b) of HC from MCI, c) of MCI converters from non-converters and d) of MCI from AD, ordered according to type of data used: T1W only, T1W+other sequences, T1W+non-imaging data, and T1W+other sequences+non-imaging data.

Finally, restricting comparisons of accuracy to studies that examined more than one diagnostic classification (Figure 3a-d), clearly demonstrates the lower accuracy for differentiating between healthy controls and MCI, or MCI from AD, or either healthy controls or AD and MCI converting/non converting, from healthy controls or AD (Figure 3 a-d).

**Healthy controls v MCI and Healthy controls v Alzheimer's**

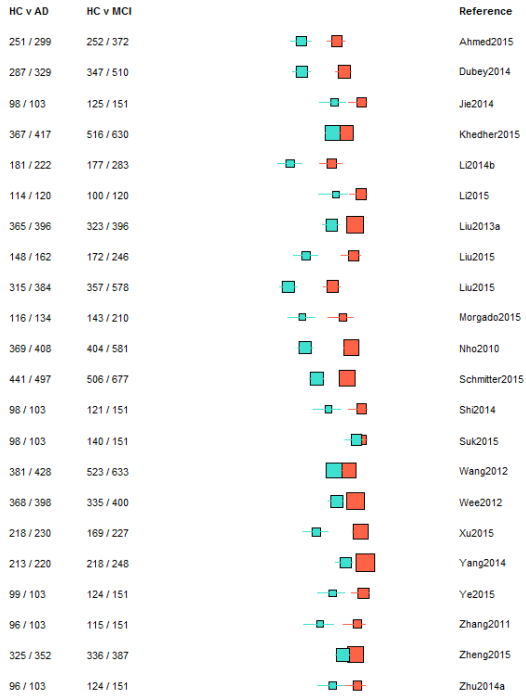

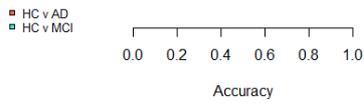

**Healthy controls v MCI and MCI converting v MCI non-converting**

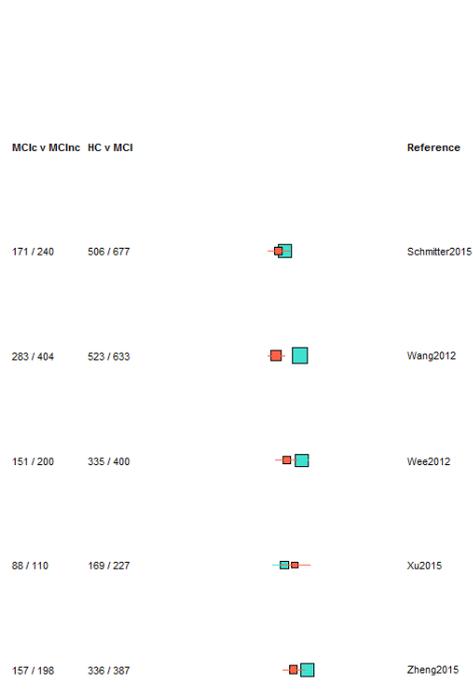

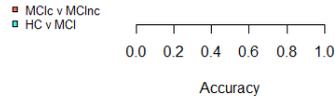

**MCI converting v MCI non-converting and MCI v Alzheimer's**

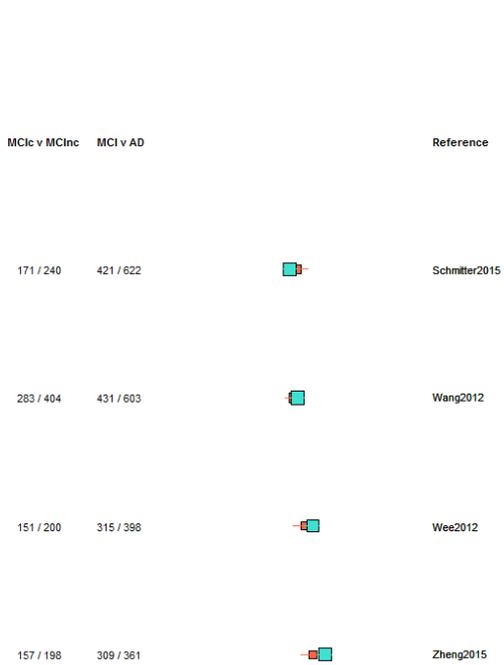

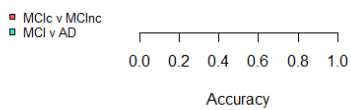

**Healthy controls v Alzheimer's and MCI v Alzheimer's**

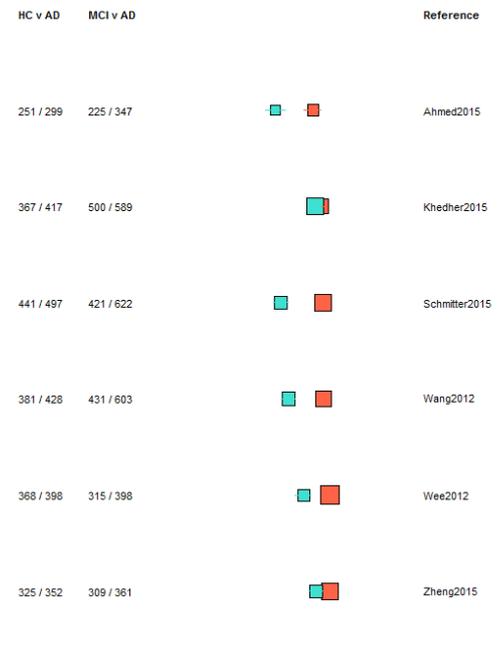

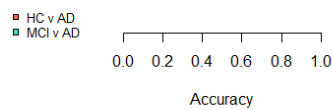

**Figure 3** Studies which included more than one diagnostic classification. Top Left. Healthy controls versus MCI and Healthy controls vs AD, Top Right. Healthy controls v MCI converting and MCI converting v MCI non-converting Bottom Left. MCI converting v MCI non-converting and MCI v AD, Bottom Right, Healthy controls v AD and MCI v AD

## DISCUSSION

We found acceptable accuracy for all machine learning methods in differentiating healthy controls from AD, but fewer data and lower accuracies for differentiating healthy controls from MCI, or MCI from AD, or (of more concern) for risk prediction of MCI non-converters from converters to AD. From a clinical perspective, the comparison of healthy controls to AD is the least important distinction: such Type I diagnostic studies do not aim to produce clinically relevant estimates of sensitivity and specificity, but to test the initial feasibility of a method. While the results for machine learning methods in differentiating healthy controls from AD are encouraging, the performance across the other cognitive diagnosis categories indicates that the field has some way to go before these methods should enter routine clinical use.[4] The over-reliance on one data source, one type of imaging, and one machine learning method, further limits the clinical relevance and generalisability of the results. This may reflect that, as yet, machine learning is still insufficiently intertwined with the clinical world, in part due to misalignment of targets and methods: while the machine learning community aims primarily for algorithm novelty, inspired largely by computer vision and machine learning, clinicians want reliable, validated, methods for early diagnosis, risk prediction, or monitoring interventions, that are better than conventional methods, and change clinical practice.

We aimed to include as many relevant papers as possible, so kept the search broad. We retained conference papers to reflect the tendency to publish conference papers that equate to full publications in the fast-moving medical image analysis, computer vision and machine learning fields. High-quality conference papers are at least as selective as many journals; e.g., MICCAI, a leading medical image analysis conference, applies a 3-stage selection protocol including rebuttal. About a quarter (29/111, 26%) of the included papers were conference papers. The number of un-refereed pre-prints becoming available online (e.g., arXiv, biorXiv) is also increasing rapidly, but we did not include these pre-print publications since they are not peer-reviewed. However, the use of these sites for dissemination is growing and may need considering in future reviews. The proportion of papers using deep learning has increased since late 2016 (including several published by the authors, many conference papers in MIUA 2018, and MICCAI 2017), and therefore this review may under-represent the most recent developments in machine learning.

However, many of these recent papers focused on methods to detect single brain lesion types, such as WMH or atrophy, that are associated with cognitive decline but not on degrees of cognitive decline itself, or on differentiating AD from healthy controls rather than more subtle diagnoses. Therefore it is unlikely that the conclusions of the present analysis, which is based on a substantial body of work, would change by the inclusion of these most recent papers.

Some non-systematic reviews and surveys on machine learning have been published.[5-11] We used established systematic review methods including QUADAS-2 criteria to grade study quality, since there are no agreed guidelines for reviews in data science and machine learning, but found the QUADAS criteria difficult to apply. We aimed to make reasonable exclusion criteria (publications from 2006 onward, data set larger than 100 for patient/image level classification, data set larger than 25 for pixel/voxel level segmentation), based on experience and consultation with a team of experts. We do not believe that the main conclusions would change significantly by including more small studies, and believe that the main messages embedded in the current literature are captured well by the review.

We excluded more than 200 papers (Supplementary Table S2) because the sample size or ground truth annotations were too small. This suggests the need for more public data repositories with annotated, reliable data. Various international initiatives provide public annotated data sets for competitions, e.g. the challenges organized by MICCAI or ISBI. Such challenges emphasize the competition aspect (achieving the best values for specific performance parameters), more than maximizing the amount of data made available, the generalisability of the results, or relevance to clinical practice. The latter two should receive more attention if the field is to advance.

We excluded many papers that did not provide accuracy data. This suggests a need to standardise reporting of performance criteria, an issue in the validation of algorithms and software for data and image analysis.[12-14] Some aspects of the perceived importance of standard criteria and data sets is highlighted by the clear majority of papers using the ADNI data set (www.adni-info.org). Although use of one dataset may promote cross-comparisons of results, it is likely to inflate estimates of accuracy and considerably reduces the generalisability of the results to clinical practice. Deep learning techniques are rapidly becoming the methods of choice in medical image analysis, and feature in increasing proportions in conferences and journals, e.g. many conference papers at MIUA 2017. However, the overall message remains the same, i.e.

differentiation of AD from healthy controls, but fewer studies and poorer accuracy at differentiating MCI vs. healthy control or AD, or MCI converters/non-converters to AD, with the same problems of sample size and repeated use of the same data and lack of clinical integration. This further increases the need for large datasets as convolutional neural networks have millions of parameters to train. The performance of systems classifying brain images as associated with AD or not seems to improve when taking into consideration multiple data types.[15, 16] Including non-imaging features, like CSF biomarkers and cognitive test scores, unsurprisingly also improve performance. Further work is needed to clarify the interplay between data from images and from other sources.[17]

Most studies started with pre-processed features ('ground truth') as input to the machine learning method. Many pre-processing techniques use population templates that derive from young populations; these are of limited relevance to the older brain and may bias the resulting machine learning outputs.[17] Very few papers on lesion segmentation techniques were included as most failed the inclusion criteria on annotations (ground truth). This reflects that generating sufficient ground truth for a reliable validation of such algorithms is very time consuming, and highlights a limitation of machine learning methods in relying on ground truth. Use of crowd-sourcing to annotate images may be one solution but would have to achieve high reliability to meet the definition of 'ground truth';[18-20] their use remains *sub judice* and depends on the application. We also notice recent work on the automatic generation of annotations (auto-annotations) for non-medical classifiers with large numbers of classes,[21] and the growing interest of medical image analysts in techniques to minimise the number of annotations required without affecting performance.[22]

It proved particularly difficult to locate papers attempting stratification of different types of dementia, and few studies combined imaging with other data types. Possible reasons include that diagnosing dementia is not a clear-cut process, so that several covariates should be considered in addition to a binary label (dementia/no dementia), e.g. time of diagnosis, source data for diagnosis (MCI test, brain images, clinical records, prescriptions). Different dementia components might be present at the same time. Finally, to our best knowledge, no public data sets exist which offer reliably stratified, sufficiently large cohorts with brain imaging.

Practically all the included papers were written for a computer science or engineering audience. They focused on technical information (e.g. algorithm choice and description, parameter setting techniques, training protocol) omitting essential clinically-relevant information (e.g. patient and cohort demographics, clinical covariates, data acquisition protocols). Clearly, specialized journals and conferences require specialist language, but international efforts are needed to make technical papers more understandable to a clinical audience, and vice versa, to improve interdisciplinarity.

## CONCLUSIONS

The results of our review indicate that machine learning methods to predict risk of dementia are not yet ready for routine use. There is a need to push inter-disciplinary collaborations, including the development of internationally agreed (by clinicians and computer science/engineers) validation protocols and clinical trials. The further development of any machine learning methods in neuroimaging requires much greater interdisciplinary working, use of varied and clinically-relevant public data sets with annotations, or ground truth, including a variety of imaging types not just T1, to maximise the use of relevant predictive variables and ensure that the resulting machine learning methods are robust and reliable prior to further testing in clinical trials in patients.

## DECLARATION OF INTEREST STATEMENT

The authors declare that the research was conducted in the absence of any commercial or financial relationships that could be construed as a potential conflict of interest.

## AUTHOR CONTRIBUTIONS

Study conception and design: EP, GM, ET, JW.

Acquisition of data: EP, VGC, MVH, LB, DA, SD, SMM, DJ, CP.

Analysis and interpretation of data: all coauthors.

Drafting of manuscript: EP, LB, TM, ET, JW.

Critical revision: EP, GM, FC, TM, ET, JW.

## FUNDING

This work was supported by the Engineering and Physical Sciences Research Council (EPSRC) grant "Multi-modal retinal biomarkers for vascular dementia" (EP/M005976/1), the Row Fogo Charitable Trust through the Row Fogo Centre for Research into Ageing and the Brain (Ref No: AD.ROW4.35. BRO-D.FID3668413), Age UK and UK Medical Research Council (G0701120, G1001245 and MR/M013111/1), the Fondation Leducq Transatlantic Network of Excellence for the Study of Perivascular Spaces in Small Vessel Disease, (ref no. 16 CVD 05), the UK Dementia Research Institute at The University of Edinburgh and the European Union Horizon 2020, PHC-03-15, project No 666881, 'SVDs@Target'. Support from NHS Lothian R&D, Edinburgh Imaging and the Edinburgh Clinical Research Facility at the University of Edinburgh is gratefully acknowledged.

# References

1.    GBD 2015 DALYs and HALE Collaborators. Global, regional, and national disability-adjusted life-years (dalys) for 315 diseases and injuries and healthy life expectancy (hale), 1990-2015: A systematic analysis for the global burden of disease study 2015. *Lancet*. 2016;388:1603-1658.

2.    Feigin VL, Forouzanfar MH, Krishnamurthi R, Mensah GA, Connor M, Bennett DA, et al. Global and regional burden of stroke during 1990-2010: Findings from the global burden of disease study 2010. *Lancet*. 2014;383:245-254.

3.    Wardlaw JM, Smith EE, Biessels GJ, Cordonnier C, Fazekas F, Frayne R, et al. Neuroimaging standards for research into small vessel disease and its contribution to ageing and neurodegeneration. *The Lancet Neurology*. 2013;12:822-838.

4.    Lancet T. Artificial intelligence in health care: Within touching distance. *Lancet*. 2017;390:2739

5.    Zheng C, Xia Y, Pan Y, Chen J. Automated identification of dementia using medical imaging: A survey from a pattern classification perspective. *Brain Inform*. 2016;3:17-27.

6.    Cure S, Abrams K, Belger M, Dell'agnello G, Happich M. Systematic literature review and meta-analysis of diagnostic test accuracy in alzheimer's disease and other dementia using autopsy as standard of truth. *J Alzheimers Dis*. 2014;42:169-182.

7.    Christian S, Petronilla B, Isabella C. Frontiers for the early diagnosis of ad by means of mri brain imaging and support vector machines. *Current Alzheimer Research*. 2016;13:509-533.

8.    Cheng B, Wee C-Y, Liu M, Zhang D, Shen D. Brain disease classification and progression using machine learning techniques. In: Suzuki K, ed. *Computational intelligence in biomedical imaging*. New York, NY: Springer New York; 2014:3-32.

9.    Shen D, Wee C-Y, Zhang D, Zhou L, Yap P-T. Machine learning techniques for ad/mci diagnosis and prognosis. In: Dua S, Acharya UR, Dua P, eds. *Machine learning in healthcare informatics*. Berlin, Heidelberg: Springer Berlin Heidelberg; 2014:147-179.

10.   Kloppel S, Abdulkadir A, Hadjidemetriou S, Issleib S, Frings L, Thanh TN, et al. A comparison of different automated methods for the detection of white matter lesions in mri data. *Neuroimage*. 2011;57:416-422.

11.   Arbabshirani MR, Plis S, Sui J, Calhoun VD. Single subject prediction of brain disorders in neuroimaging: Promises and pitfalls. *NeuroImage*. 2017;145:137-165.

12.   Trucco E, Ruggeri A, Karnowski T, Giancardo L, Chaum E, Hubschman JP, et al. Validating retinal fundus image analysis algorithms: Issues and a proposal. *Investigative ophthalmology & visual science.* 2013;54:3546-3559.

13.   Maier-Hein L, Groch A, Bartoli A, Bodenstedt S, Boissonnat G, Chang PL, et al. Comparative validation of single-shot optical techniques for laparoscopic 3-d surface reconstruction. *IEEE Trans Med Imaging.* 2014;33:1913-1930.

14.   Jannin P, Grova C, Maurer CR. Model for defining and reporting reference-based validation protocols in medical image processing. *International Journal of Computer Assisted Radiology and Surgery.* 2006;1:63-73.

15.   Li Y, Yan J, Wang P, Lv Y, Qiu M, he X. Classification of alzheimer's disease based on multiple anatomical structures' asymmetric magnetic resonance imaging feature selection. *Neural Information Processing.* 2015:280-289.

16.   Liu M, Zhang D, Adeli E, Shen D. Inherent structure-based multiview learning with multitemplate feature representation for alzheimer's disease diagnosis. *IEEE Transactions on Biomedical Engineering.* 2016;63:1473-1482.

17.   BRAINS (Brain Imaging in Normal Subjects) Expert Working Group, Shenkin SD, Pernet C, Nichols TE, Poline JB, Matthews PM, et al. Improving data availability for brain image biobanking in healthy subjects: Practice-based suggestions from an international multidisciplinary working group. *Neuroimage.* 2017;153:399-409.

18.   Mitry D, Zutis K, Dhillon B, Peto T, Hayat S, Khaw KT, et al. The accuracy and reliability of crowdsource annotations of digital retinal images. *Translational vision science & technology.* 2016;5:6.

19.   Albarqouni S, Baur C, Achilles F, Belagiannis V, Demirci S, Navab N. Aggnet: Deep learning from crowds for mitosis detection in breast cancer histology images. *IEEE Transactions on Medical Imaging.* 2016;35:1313-1321

20.   Adriana K, Olga R, Li F-F, Kristen G. *Crowdsourcing in computer vision.* Now Foundations and Trends; 2016.

21.   Guillaumin M, Küttel D, Ferrari V. Imagenet auto-annotation with segmentation propagation. *International Journal of Computer Vision.* 2014;110:328-348.

22.   Valindria VV, Lavdas I, Bai W, Kamnitsas K, Aboagye EO, Rockall AG, et al. Reverse classification accuracy: Predicting segmentation performance in the absence of ground truth. *IEEE Transactions on Medical Imaging.* 2017;36:1597-1606.

**Supplementary Methods, Tables and Figures**

**Methods**

Calculation of the precision and recall for lesion identification tasks
Precision and Recall are defined as:

$$\text{Precision} = \frac{tp}{tp + fp}$$

$$\text{Recall} = \frac{tp}{tp + fn}$$

where:  tp= true positive (indicates accurate lesion identification)
tn=true negative (indicates correct rejection of non-lesion tissue)
fp=false positive (indicates identification of a lesion that is not there)
fn=false negative (indicates failure to identify a lesion that is present)

**Table S1**. Acronyms used in Tables listing study results.

| Acronym | Definition |
| --- | --- |
| AD | Alzheimer's Disease |
| AdaBoost | Adaptive Boosting |
| ADAS-Cog | Alzheimer's Disease Assessment Scale - cognitive subtest |
| ADNI | Alzheimer's Disease Neuroimaging Initiative |
| APOE3 | Apolipoprotein E |
| BoW | Bag of Words |
| CCA | Canonical Correlation Analysis |
| CDR-SB | Clinical Dementia Rating - Sum of Boxes |
| CMB | Cerebral Microbleeds |
| CSF | Cerebrospinal Fluid |
| DLB | Dementia with Lewy Bodies |
| DTI | Diffusion Tensor Imaging |
| DWT | Discrete Wavelet Transform |
| FA | Fractional Anisotropy |
| FAQ | Functional Activities Questionnaire |
| FDG | Fluorodeoxyglucose |
| FDR | Fisher Discriminant Ratio |
| FF-NN | Feed Forward Neural Network |
| FLAIR | Fluid Attenuated Inversion Recovery |
| fMRI | functional Magnetic Resonance Imaging |
| FS | Feature Selection |
| FTD | Frontotemporal Dementia |
| GM | Gray Matter |
| GRE | Gradient Recalled Echo |
| HC | Healthy Control |
| HMM | Hidden Markov Model |
| ICV | Intracranial Volume |
| kNN | k-Nearest Neighbours |
| LASSO | Least Absolute Shrinkage and Selection Operator |
| LBP | Local Binary Patterns |
| LDA | Linear Discriminant Analysis |
| MABMIS | Multi-Atlas based Multi-Image Segmentation |
| MCI | Mild Cognitive Impairment |
| MCIc | Mild Cognitive Impairment (converting) |
| MCIe | Mild Cognitive Impairment (early amnestic) |
| MCInc | Mild Cognitive Impairment (non-converting) |
| MD | Mean Diffusivity |
| MIL | Multiple Instance Learning |
| MKL | Multiple Kernel Learning |
| MMSE | Mini-Mental State Examination |
| mRMR | minimum Redundancy Maximum Relevance |
| MTI | Magnetization Transfer Imaging |
| NN | Neural Network |
| OASIS | Open Access Series of Imaging Studies |
| OPLS | Orthogonal Projections to Latent Structures |
| PCA | Principal Component Analysis |
| PD | Proton Density |
| PDF | Probability Distribution Function |
| PESFAM | Probabilistic Ensemble Simplified Fuzzy ARTMAP |
| PET | Positron Emission Tomography |
| P-NN | Probabilistic Neural Network |
| QDA | Quadratic Discriminant Analysis |
| QDC | Quadratic Discriminant Classifier |
| RAVENS | Regional Analysis of Volumes Examined in Normalized Space |
| RAVLT | Rey's Auditory Verbal Learning Test |
| RBF | Radial Basis Function |
| RFE | Recursive Feature Elimination |
| ROI | Region Of Interest |
| SAE | Stacked Auto Encoders |
| SES | Socioeconomic Status |
| SRC | Sparse Representation Classification |
| SVD | Small Vessel Disease |
| SVM | Support Vector Machine |
| SWI | Susceptibility-Weighted Imaging |
| TIA | Transient Ischemic Attack |
| VBM | Voxel-Based Morphometry |
| WM | White Matter |
| WML | White Matter Lesions |

**Table S2**. Details of reason for rejection and proportions

| Rejection criteria | Number of rejected studies (%) |
|---|---|
| 1.  Animals or ex-vivo. | 1 (0.2) |
| 2.  Review, survey, collection. | 27 (5.3) |
| 3.  Size of the dataset, number of observers. | 209 (40.7) |
| 4.  Semi-automatic technique. | 61 (11.9) |
| 5. No structural imaging. | 25 (4.9) |
| 6. Pre-processing technique. | 12 (2.3) |
| 7. Healthy region parcellation. | 13 (2.5) |
| 8. Non-comparable results | 96 (18.7) |
| 9. "Multiple" publications. | 63 (12.3) |
| 10. Non-reproducible | 7 (1.4) |

## Table S3. Machine learning studies that classified AD or MCI and healthy controls

| Reference | Dataset | Classification tasks (n) | Image features (FS and representation) | Additional imaging sequences and features | Classifiers | Results |
|---|---|---|---|---|---|---|
| (Aggarwal, Rana et al. 2015) | OASIS | HC vs AD (99 / 99) | 3D-DWT (symmlet) of 7 ROI's: hippocampus, amygdalae, ventricles, anterior and posterior cingulate (FS by FDR and mRMR). | n.a. | kNN | Sen = 0.789 / Spe = 0.810 |
| (Aguilar, Westman et al. 2013) | AddNeuroMed | HC vs AD (110 / 116)<br>MCInc vs MCIc (98 / 21) | 68 cortical thickness values and 50 regional volumes obtained with FreeSurfer. | Education.<br>n.a. | SVM (non-lin)<br>OPLS | Sen = 0.862 / Spe = 0.900<br>Sen = 0.810 / Spe = 0.684 |
| (Ahmed, Mizotin et al. 2015) | ADNI | HC vs AD (162 / 137)<br>HC vs MCI (162 / 210)<br>AD vs MCI (210 / 137) | Circular harmonic features extracted from hippocampus and posterior cingulate cortex (FS by PCA; BoW representation) | n.a. | SVM (RBF) | Sen = 0.791 / Spe = 0.882<br>Sen = 0.626 / Spe = 0.748<br>Sen = 0.490 / Spe = 0.752 |
| (Anagnostopoulos, Giannoukos et al. 2013) | AddNeuroMed | HC vs AD (113 / 122 / 123) | Cortical volume and thickness for specific ROI's, manual volume measurement of the hippocampus. | T2w, demographics. | Ensemble of FF-NN, SVM, PESFAM, P-NN, kNN | Acc = 0.771 |
| (Archana and Ramakrishnan 2014) | OASIS | HC vs AD (92 / 45)<br>HC vs MCI (92 / 67)<br>AD vs MCI (67 / 45) | Voxel-wise texture features from structure tensor analysis (FS by FDR). | n.a. | SVM | Sen = 0.877 / Spe = 0.849<br>Sen = 0.764 / Spe = 0.783<br>Sen = 0.747 / Spe = 0.767 |
| (Babu, Suresh et al. 2013) | ADNI | HC vs MCIc (232 / 167)<br>MCInc vs MCIc (236 / 167) | Voxel-wise GM probability values from VBM analysis (FS by t-test). | n.a. | RBF-NN | Sen = 0.730 / Spe = 0.840<br>Sen = 0.880 / Spe = 0.890 |
| (Beheshti, Demirel et al. 2015) | ADNI | HC vs AD (130 / 130) | Voxel-wise GM probability values from VBM analysis (FS based on PDF of ROI's). | n.a. | SVM (RBF) | Sen = 0.908 / Spe = 0.908 |
| (Casanova, Hsu et al. 2013) | ADNI | HC vs AD (188 / 171)<br>HC vs MCInc (188 / 182)<br>HC vs MCIc (188 / 153)<br>MCInc vs MCIc (182 / 153) | Voxel-wise intensities from GM, WM and CSF maps. | n.a.<br>n.a.<br>n.a.<br>Cognitive scores. | Regularized logistic regression | Sen = 0.843 / Spe = 0.890<br>Sen = 0.586 / Spe = 0.681<br>Sen = 0.740 / Spe = 0.831<br>Sen = 0.579 / Spe = 0.701 |
| (Chaddad, Desrosiers et al. 2016) | OASIS | HC vs AD (62 / 62) | 3D co-occurrence matrix. | n.a. | Random forest. | Sen = 0.742 / Spe = 0.759 |
| (Chen and Pham 2013) | OASIS | HC vs AD (75 / 75) | 2D regularity information from semi-variogram analysis of GM maps. | n.a. | HMM | Sen = 0.800 / Spe = 0.800 |
| (Chen, Wei et al. 2015) | ADNI | MCInc vs MCIc (167 / 236) | GM volumes in 93 ROI's (sparse representation). | n.a. | SRC | Sen = 0.581 / Spe = 0.763 |
| (Chincarini, Bosco et al. 2011) | ADNI | HC vs AD (189 / 144)<br>HC vs MCInc (189 / 136)<br>MCInc vs MCIc (166 / 136) | Voxel intensities of filtered masks in 9 ROI's: hippocampi, amygdalae. middle and inf temp gyri, rolandic. | n.a. | Random forest + SVM | Sen = 0.890 / Spe = 0.940<br>Sen = 0.890 / Spe = 0.800<br>Sen = 0.720 / Spe = 0.650 |
| (Cho, Seong et al. 2012) | ADNI | HC vs AD (160 / 66)<br>HC vs MCInc (80 / 35)<br>HC vs MCIc (66 / 35) | Cortical thickness values (FS by PCA). | n.a. | LDA | Sen = 0.820 / Spe = 0.930<br>Sen = 0.660 / Spe = 0.890<br>Sen = 0.630 / Spe = 0.760 |
| (Costafreda, Dinov et al. 2011) | AddNeuroMed | MCInc vs MCIc (81 / 22) | Thickness values of hippocampi. | n.a. | SVM (RBF) | Sen = 0.770 / Spe = 0.800 |
| (Coupé, Fonov et al. 2015) | ADNI | MCInc vs MCIc (309 / 37) | SNIPE (Scoring by Nonlocal Image Patch Estimator) hippocampal features. | n.a. | LDA | Sen = 0.649 / Spe = 0.735 |
| (Cui, Wen et al. 2012) | Local | HC vs MCI (204 / 79) | 10 regional volumes from T1w, 58 WM integrity features from DTI. | DTI. | SVM (RBF) | Sen = 0.520 / Spe = 0.784 |
| (Cuingnet, Gerardin et al. 2011) | ADNI | HC vs AD (80 / 66)<br>HC vs MCIc (80 / 35)<br>MCInc vs MCIc (66 / 35) | Voxel-wise GM probability values in ROI's defined by different processing pipelines. | n.a. | SVM (linear) | en = 0.810 / Spe = 0.950<br>Sen = 0.680 / Spe = 0.950<br>Sen = 0.570 / Spe = 0.780 |
| (Cuingnet, Glaunes et al. 2013) | ADNI | HC vs AD (81 / 68) | GM, WM and CSF probability maps, cortical thickness values (FS by anatomical and spatial priors in SVM). | n.a. | SVM (non-lin) | Sen = 0.800 / Spe = 0.930 |
| (Davazikos, Bhatt et al. 2011) | ADNI | MCInc vs MCIc (85 / 35) | Pattern of atrophy in GM and WM maps. | CSF biomarkers. | SVM (non-lin) | Sen = 0.842 / Spe = 0.512 |
| (Ding, Zhang et al. 2015) | ADNI | HC vs AD (58 / 54) | 8 GM volumes in ROI's, 220 texture features, 64 features from 2D multiscale Gabor filtering (FS by RFE). | n.a. | SVM | Sen = 0.870 / Spe = 0.983 |
| (Dubey, Zhou et al. 2014) | ADNI | HC vs AD (191 / 138)<br>HC vs MCI (191 / 319)<br>HC vs AD+MCIc (191 / 280) | Cortical thickness values, volumes of cortical ROI's, volumes of WM in ROI's, total surface area of the cortex (FS by sparse logistic regression). | n.a. | Random forest<br>SVM<br>SVM | Sen = 0.826 / Spe = 0.906<br>Sen = 0.793 / Spe = 0.493<br>Sen = 0.879 / Spe = 0.825 |
| (Dyrba, Ewers et al. 2012) | Local | HC vs AD (143 / 137) | GM probability map values (FS by entropy-based information gain). | n.a. | SVM (RBF) | Sen = 0.874 / Spe = 0.912 |
| (Eskildsen, Coupé et al. 2013) | ADNI | HC vs AD (226 / 194)<br>MCInc vs MCIc (227 / 161) | Cortical thickness values (FS by t-test and mRMR). | Age. | LDA | Sen = 0.794 / Spe = 0.889<br>Sen = 0.658 / Spe = 0.683 |
| (Eskildsen, Coupé et al. 2015) | ADNI | MCInc vs MCIc (238 / 167) | L / R hippocampal grading, cortical thickness values of 3 ROI's (FS by mutual information method). | Age. | LDA | Sen = 0.696 / Spe = 0.736 |
| (Filipovych, Davatzikos et al. 2011) | ADNI | HC vs AD (63 / 54)<br>MCInc vs MCIc (174 / 68) | GM RAVENS map (FS by RFE). | n.a. | SVM (linear, semi-supervised) | Sen = 0.796 / Spe = 0.857<br>Sen = 0.794 / Spe = 0.517 |
| (Granziera, Daducci et al. 2015) | Local | HC vs MCI (77 / 42) | Volume and mean intensity values from 7 ROI's (WM and cortical GM, thalamus, caudate, globus pallidus, putamen and hippocampus) and for WM and GM of each lobe. | MTI, T2*. | SVM | Sen = 0.600 / Spe = 0.830 |
| (Gray, Aljabar et al. 2013) | ADNI | HC vs MCI (35 / 75) | 83 ROI volumes from GM maps, voxel intensities from PET. | PET, APOE3, CSF biomarkers. | Random forest | Sen = 0.775 / Spe = 0.679 |
| (Guerrero, Wolz et al. 2014) | ADNI<br>ADNI-GO | HC vs AD (175 / 106)<br>MCInc vs MCIc (114 / 116)<br>HC vs MCIc (175 / 116)<br>HC vs MCInc (134 / 229) | Voxel intensities (Laplacian eigenmaps representation after FS by Elastic Net and manifold learning). | n.a. | SVM (linear) | Sen = 0.860 / Spe = 0.850<br>Sen = 0.750 / Spe = 0.670<br>Sen = 0.860 / Spe = 0.760<br>Sen = 0.610 / Spe = 0.690 |
| (Herrera, Rojas et al. 2013) | ADNI | HC vs AD (443 / 459)<br>HC vs MCI vs AD (443 / 448 / 459) | 2D-DWT (Db4 and Haar) multi-scale features (FS by PCA and mutual information method). | n.a. | SVM (RBF) | Sen = 0.963 / Spe = 0.961<br>Acc = 0.774 |
| (Hinrichs, Singh et al. 2009) | ADNI | HC vs AD (94 / 89) | GM probability maps (FS by t-test to select relevant voxels). | n.a. | Linear programming | Sen = 0.850 / Spe = 0.800 |
| (Hinrichs, Singh et al. 2011) | ADNI | HC vs AD (66 / 48) | GM probability maps (FS by t-test to select relevant voxels). | PET, APOE3, CSF biomarkers, cognitive scores. | MKL | Sen = 0.867 / Spe = 0.966 |
| (Hor and Moradi 2016) | ADNI | HC vs MCIc (178 / 96)<br>HC vs MCInc (178 / 126)<br>HC vs MCIc (118 / 144)<br>MCInc vs MCIc (126 / 96)<br>MCInc vs MCIc (144 / 27) | Volume measurements of six ROI's (ventricles, hippocampus, whole-brain, entorhinal, fusiform and mid-temporal) and ICV from T1w, FDG and AV45 uptake values from PET (FS by information gain). | n.a.<br>n.a.<br>n.a.<br>n.a.<br>PET. | Random forest | Sen = 0.691 / Spe = 0.844<br>Sen = 0.734 / Spe = 0.863<br>Sen = 0.737 / Spe = 0.897<br>Sen = 0.819 / Spe = 0.750<br>Sen = 0.831 / Spe = 0.803 |
| (Hu, Wang et al. 2016) | ADNI | HC vs AD (228 / 188)<br>MCInc vs MCIc (62 / 167) | 3D-DWT (Gabor and Haar) multi-scale features from GM of map of hippocampus. | n.a. | SVM (linear) | Sen = 0.846 / Spe = 0.855<br>Sen = 0.718 / Spe = 0.823 |
| (Illan, Garriz et al. 2014) | ADNI | HC vs AD (76 / 63)<br>HC vs MCIc (76 / 110) | Binary values of GM, WM maps in 6 ROI's: parahippocampal gyrus, lingual gyrus, hippocampus, frontal lobe, precentral gyrus, temporal lobe (Bayesian network representation). | n.a. | SVM (ensemble) | Sen = 0.926 / Spe = 0.845<br>Sen = 0.773 / Spe = 0.845 |

| Reference | Dataset | Classification tasks (n) | Image features (FS and representation) | Additional imaging sequences and features | Classifiers | Results |
|---|---|---|---|---|---|---|
| (Jiang and Shi 2014) | ADNI | HC vs AD (52 / 51) | 83 ROI volumes from GM maps (FS by sparse kernel entropy component analysis). | n.a. | kNN | Sen = 0.920 / Spe = 0.904 |
| (Jie, Zhang et al. 2014) | ADNI | HC vs AD (52 / 51)<br>HC vs MCI (52 / 99) | GM volumes and PET intensity values in 93 ROI's(FS by manifold regularized multitask learning method). | PET,<br>CSF biomarkers. | MKL | Sen = 0.947 / Spe = 0.958<br>Sen = 0.894 / Spe = 0.708 |
| (Khedher, Ramirez et al. 2015) | ADNI | HC vs AD (229 / 188)<br>HC vs MCI (229 / 401)<br>MCI vs AD (401 / 188) | Voxel intensities in GM and WM maps (FS by partial least square). | n.a. | SVM (linear) | Sen = 0.913 / Spe = 0.851<br>Sen = 0.822 / Spe = 0.816<br>Sen = 0.870 / Spe = 0.838 |
| (Komlagan, Ta et al. 2014) | ADNI | MCInc vs MCIc (236 / 166) | SNIPE (Scoring by Nonlocal Image Patch Estimator) hippocampal features (FS by sparse logistic regression). | n.a. | SVM (linear) | Sen = 0.615 / Spe = 0.856 |
| (Korolev, Symonds et al. 2016) | ADNI | MCInc vs MCIc (139 / 120) | Cortical thickness, volumes, curvature and surface area of 180 ROI's (FS by joint mutual information criterion). | Risk factors, cognitive scores, proteomic data. | MKL | Sen = 0.834 / Spe = 0.764 |
| (Krashenyi, Ramirez et al. 2016) | ADNI | HC vs AD (229 / 188)<br>HC vs MCI (229 / 401)<br>MCI vs AD (401 / 188) | Mean ROI intensity values of GM and WM maps from T1w and mean intensity from PET (FS by t-test). | PET. | Fuzzy inference system. | Sen = 0.933 / Spe = 0.922<br>Sen = 0.759 / Spe = 0.861<br>Sen = 0.749 / Spe = 0.820 |
| (Lebedev, Westman et al. 2014) | ADNI | HC vs AD (75 / 35) | Volumes from 41 ROI's and cortical thickness values (FS by PCA and RFE). | APOE3, demographics. | Random forest. | Sen = 0.920 / Spe = 0.886<br>Sen = 0.833 / Spe = 0.813 |
| (Li, Wang et al. 2010) | OASIS | HC vs MCI (80 / 89) | GM values in 19 ROI's (FS by t-test and feature ranking). | MMSE. | SVM (RBF) | Sen = 0.919 / Spe = 0.880 |
| (Li, Liu et al. 2014) | ADNI | MCIc vs MCInc (161 / 132) | Cortical thickness values, volumes of cortical ROI's, volumes of WM in ROI's, total surface area of the cortex (FS by hierarchical Lasso method). | Demographics, genetic data, cognitive scores, lab tests. | Random forest | Sen = 0.667 / Spe = 0.814 |
| (Li, Oishi et al. 2014) | ADNI | HC vs AD (142 / 80)<br>MCInc vs MCIc (142 / 141) | Voxel-wise combination of 2D-LBP from axial, coronal and sagittal orientations (FS by t-test and a priori knowledge). | n.a. | SVM | Sen = 0.804 / Spe = 0.827<br>Sen = 0.615 / Spe = 0.635 |
| (Li, Yan et al. 2015) | ADNI | HC vs AD (60 / 60)<br>HC vs MCI (60 / 60) | Volume and 15 texture features from 4 structures (GM, WM, CSF, hippocampus) in L / R hemispheres (FS by chain-like agent genetic algorithm). | n.a. | SVM (RBF) | Sen = 0.927 / Spe = 0.973<br>Sen = 0.804 / Spe = 0.864 |
| (Liu, Suk et al. 2013) | ADNI | HC vs AD (198 / 198)<br>HC vs MCI (198 / 198) | Volume and cortical thickness from 68 ROI's (sparse representation and high-order graph matching for FS). | n.a. | SVM (multi-kernel) | Sen = 0.894 / Spe = 0.950<br>Sen = 0.778 / Spe = 0.854 |
| (Liu, Tosun et al. 2013) | ADNI | HC vs MCIc (138 / 97)<br>HC vs MCInc (138 / 93)<br>HC vs AD (138 / 86)<br>MCInc vs MCIc (93 / 97)<br>MCInc vs AD (93 / 86)<br>MCIc vs AD (97 / 86) | Values of volume from 94 ROI's and cortical thickness from 68 ROI's (representation by local linear embedding FS by Elastic Net). | n.a. | Logistic regression | Sen = 0.650 / Spe = 0.630<br>Sen = 0.810 / Spe = 0.820<br>Sen = 0.860 / Spe = 0.930<br>Sen = 0.800 / Spe = 0.560<br>Sen = 0.770 / Spe = 0.730<br>Sen = 0.560 / Spe = 0.610 |
| (Liu, Zhang et al. 2013) | ADNI | HC vs AD (229 / 198)<br>HC vs MCInc (229 / 236)<br>MCInc vs AD (236 / 167) | Voxel-wise GM probability values (regularized tree-structured approach for sparse learning). | n.a. | SVM (linear) | Sen = 0.853 / Spe = 0.943<br>Sen = 0.801 / Spe = 0.922<br>Sen = 0.562 / Spe = 0.809 |
| (Liu, Zhou et al. 2014) | ADNI | HC vs AD (70 / 50) | 126 hippocampal shape features and GM volumes from 100 ROI's/ (FS by LASSO). | CSF biomarkers. | MKL | Sen = 0.933 / Spe = 0.875 |
| (Liu, Liu et al. 2015) | ADNI | HC vs AD (204 / 180)<br>HC vs MCI (204 / 374)<br>HC vs MCInc vs AD (204 / 214 / 160 / 180)<br>HC vs AD (77 / 85)<br>HC vs MCI (77 / 169)<br>HC vs MCInc vs MCIc vs AD (77 / 102 / 67 / 85) | Values of volume from T1w and of metabolic rate of glucose consumption from PET in 83 ROI's (FS by Elastic Net). | n.a.<br>n.a.<br>n.a.<br><br>PET<br>PET<br>PET | Deep learning (SAE) | Sen = 0.868 / Spe = 0.778<br>Sen = 0.495 / Spe = 0.843<br>Acc = 0.463<br><br>Sen = 0.923 / Spe = 0.904<br>Sen = 0.600 / Spe = 0.923<br>Acc = 0.538 |
| (Liu, Cai et al. 2016) | ADNI | HC vs MCI vs AD (77 / 169 / 85) | GM Volume, local gyrification index, convexity and solidity ratios from T1w, mean index, fuzzy index, 3 difference-of-Gaussian features from PET in 83 ROI's. | PET. | SVM (multi-kernel) | ACC = 0.6535 |
| (Liu, Zhang et al. 2016) | ADNI | HC vs AD (128 / 97)<br>MCInc vs MCInc (117 / 117) | ROI average values from different templates of GM density maps (FS and encoding by subclass clustering). | n.a. | SVM (ensemble) | Sen = 0.928 / Spe = 0.957<br>Sen = 0.860 / Spe = 0.784 |
| (Luchtenberg, Simões et al. 2014) | OASIS | HC vs MCI (66 / 70) | Dissimilarity matrix of voxel intensity histograms. | n.a. | kNN | Sen = 0.800 / Spe = 0.880 |
| (Martinez-Torteya, Treviño et al. 2015) | ADNI | MCI vs AD (86 / 24) | GM volume in 90 ROI's, cortical thickness in 139 ROI's from T1w, metabolic rate of glucose consumption from PET (FS by genetic models and Pearson correlation coefficients). | PET, CSF biomarkers, APOE3, plasma, biological samples. | LDA | Sen = 0.476 / Spe = 0.941 |
| (Martinez-Murcia, Gorriz et al. 2016) | ADNI | HC vs AD (180 / 180) | LBP features from GM and WM maps (2D representation of the brain and FS by t-test). | n.a. | SVM (linear) | Sen = 0.899 / Spe = 0.919 |
| (McEvoy, Fennema-Notestine et al. 2009) | ADNI | HC vs AD (139 / 84) | Morphometric measures from 58 ROI's. | Age, RAVLT, ADAS-Cog, MMSE, CDR-SB, FAQ. | LDA | Sen = 0.830 / Spe = 0.930 |
| (Moradi, Pepe et al. 2015) | ADNI | MCInc vs MCIc (100 / 164) | Voxel-wise GM density values (FS by regularized logistic regression). | | Random forest + SVM (RBF) | Sen = 0.870 / Spe = 0.740 |
| (Morgado and Silveira 2015) | ADNI | HC vs AD (75 / 59)<br>HC vs MCI (75 / 135) | Voxel-wise GM density values (FS by Minimal Neighbourhood Redundancy Maximal Relevance). | n.a. | SVM | Sen = 0.869 / Spe = 0.872<br>Sen = 0.688 / Spe = 0.670 |
| (Nho, Shen et al. 2010) | ADNI | HC vs AD (226 / 182)<br>HC vs MCI (226 / 355) | GM density values from 86 ROI's, cortical thickness values from 56 ROI's (FS by SVM-RFE). | APOE3, family history. | SVM (RBF) | Sen = 0.850 / Spe = 0.948<br>Sen = 0.694 / Spe = 0.698 |
| (Plocharski and Østergaard 2016) | ADNI | HC vs AD (96 / 109) | Depth, length, curvature and surface area of 24 sulci (FS by forward selection). | n.a. | SVM (linear) | Sen = 0.900 / Spe = 0.867 |
| (Rao, Lee et al. 2011) | Local | HC vs AD (60 / 69) | Voxel-wise GM density values (FS by spatially regularized formulation). | n.a. | Logistic regression | Sen = 0.904 / Spe = 0.803 |
| (Rueda, Gonzalez et al. 2014) | OASIS | HC vs MCI + AD (98 / 100) | Voxel intensities (graph-based saliency map representation). | n.a. | MKL | Sen = 0.670 / Spe = 0.735 |

| Reference | Dataset | Classification tasks (n) | Image features (FS and representation) | Additional imaging sequences and features | Classifiers | Results |
|---|---|---|---|---|---|---|
| (Savio and Graña 2013) | OASIS | HC vs AD (316 / 100) | Voxel intensities (represented as the trace of the Jacobian matrix from tensor-based morphometry analysis. FS by t-test). | n.a. | SVM (RBF) | Sen = 0.856 / Spe = 0.863 |
| (Schmitter, Roche et al. 2015) | ADNI | HC vs AD (276 / 221) HC vs MCI (276 / 401) MCI vs AD (401 / 221) MCInc vs MCIc (103 / 137) | Volumes, obtained from FreeSurfer or MorphoBox of: total GM, left and right temporal GM, total CSF, and lateral, 3 and 4 ventricles. | n.a. | SVM | Sen = 0.860 / Spe = 0.910 Sen = 0.690 / Spe = 0.830 Sen = 0.690 / Spe = 0.670 Sen = 0.750 / Spe = 0.660 |
| (Schouten, Koini et al. 2016) | Local | HC vs AD (173 / 77) HC vs AD mild (173 / 39) HC vs AD moderate (173 / 38) | GM density values in 110 ROI's, WM density values in 20 ROI's from T1w. FA and MD values in 20 ROI's. 2415 values of correlation from functional connectivity analysis from fMRI (FS by Elastic Net). | DTI, fMRI. | Regularized logistic regression | Sen = 0.826 / Spe = 0.927 Sen = 0.721 / Spe = 0.935 Sen = 0.813 / Spe = 0.956 |
| (Shi, Suk et al. 2014) | ADNI | HC vs AD (52 / 51) HC vs MCI (52 / 99) | GM volumes and PET intensity values in 93 ROI's (FS by Lasso). | PET | SVM (linear) SRC | Sen = 0.942 / Spe = 0.969 Sen = 0.817 / Spe = 0.762 |
| (Singh, Fletcher et al. 2014) | ADNI | MCInc vs MCIc (73 / 54) | Anatomical shape variations with respect to atlas (FS by partial least square model). | PET, APOE3, CSF biomarkers. | QDA | Sen = 0.942 / Spe = 0.969 Sen = 0.817 / Spe = 0.762 |
| (Spulber, Simmons et al. 2013) | AddNeuroMed, ADNI | HC vs AD (52 / 51) HC vs MCI (52 / 99) | Volumes of 23 ROI's and cortical thickness values of 34 ROI's. | PET | OPLS | Sen = 0.861 / Spe = 0.904 Sen = 0.696 / Spe = 0.668 |
| (Suk, Lee et al. 2015) | ADNI | HC vs AD (52 / 51) HC vs MCI (52 / 99) HC vs MCI vs AD(229 / 403 / 198) HC vs MCInc vs MCIc vs AD (229 / 236 / 167 / 198) | GM volumes and PET intensity values in 93 ROI's (FS by deep weighted subclass-based sparse multi-task learning approach). | PET, CSF biomarkers. | SVM | Sen = 0.920 / Spe = 0.980 Sen = 0.939 / Spe = 0.908 Acc = 0.577  Acc = 0.478 |
| (Tong, Wolz et al. 2014) | ADNI | HC vs AD (231 / 198) HC vs MCIc (231 / 167) MCInc vs MCIc (238 / 167) | Voxel intensity from a variable number K of patches (MIL approach, FS by Elastic Net). | n.a. | SVM | Sen = 0.849 / Spe = 0.926 Sen = 0.689 / Spe = 0.931 Sen = 0.665 / Spe = 0.731 |
| (Varol, Gaonkar et al. 2012) | ADNI | HC vs AD (148 / 116) | Voxel-wise density values from GM, WM and ventricles maps (FS by t-test). | n.a. | SVM (ensemble) | Sen = 0.862 / Spe = 0.897 |
| (Vemuri, Gunter et al. 2008) | ADNI | HC vs AD (50 / 50) | Voxel-wise density values from GM, WM and CSF maps (FS by feature ranking). | Demographics, APOE3. | SVM (linear) | Sen = 0.860 / Spe = 0.920 |
| (Wachinger and Reuter 2016) | ADNI | HC vs MCI vs AD (129 / 122 / 103) | Cortical thickness values in 70 ROI's, volumes of 39 ROI's and 58 shape features. (FS by Elastic Net). | n.a. | Multinomial regression | Acc = 0.590 |
| (Wang, Jia et al. 2012) | ADNI | HC vs AD (229 / 199) MCI vs AD (404 / 199) HC vs MCI (229 / 404) MCInc vs MCIc (236 / 168) | GM, WM CSF values in 54 ROI's obtained from MABMIS pipeline (FS by t-test). | n.a. | SVM (linear) | Sen = 0.861 / Spe = 0.918 Sen = 0.403 / Spe = 0.870 Sen = 0.794 / Spe = 0.881 Sen = 0.667 / Spe = 0.723 |
| (Wang, Du et al. 2015) | ADNI | HC vs MCI (52 / 99) | GM volumes and PET intensity values in 93 ROI's (FS by PCA). | PET, CSF biomarkers. | SVM (linear) | Sen = 0.827 / Spe = 0.473 |
| (Wee, Yap et al. 2012) | ADNI | HC vs AD (200 / 198) HC vs MCI (200 / 200) MCI vs AD (200 / 198) MCInc vs MCIc (111 / 89) | Cortical thickness, GM and WM volumes in 68 ROI's. Correlative features between pairs of ROI's (FS by t-test, mRMR and SVM-RFE). | n.a. | SVM (RBF, multi-kernel) | Sen = 0.904 / Spe = 0.943 Sen = 0.836 / Spe = 0.840 Sen = 0.780 / Spe = 0.805 Sen = 0.635 / Spe = 0.844 |
| (Wei, Li et al. 2016) | ADNI | HC vs AD (83 / 76) | Cortical thickness, volume, and cortical surface area in 68 ROI's. 136 nodal features from the thickness network(FS by regularized sparse linear regression). | n.a. | SVM (RBF) | Sen = 0.684 / Spe = 0.759 |
| (Westman, Simmons et al. 2011) | AddNeuroMed, ADNI | HC vs AD (335 / 295) MCInc vs MCIc (353 / 84) | Cortical thickness in 57 selected ROI's and volumes of 23 ROI's. | n.a. | OPLS | Sen = 0.834 / Spe = 0.878 Sen = 0.714 / Spe = 0.601 |
| (Wolz, Julkunen et al. 2011) | AddNeuroMed, ADNI | HC vs AD (231 / 198) HC vs MCIc (231 / 167) MCInc vs MCIc (238 / 167) | Hippocampal volume, cortical thickness from different ROI's. 84 tensor-based morphometry and 20 manifold learning features (FS by t-test). | n.a. | LDA | Sen = 0.930 / Spe = 0.850 Sen = 0.860 / Spe = 0.820 Sen = 0.670 / Spe = 0.690 |
| (Xie, Cui et al. 2015) | Local | HC vs MCI (64 / 64) | Voxel-wise value of GM from T1w and FA and MD from DTI (FS by t-test). | DTI. | SVM (linear, ensemble) | Sen = 0.786 / Spe = 0.888 |
| (Xu, Wu et al. 2015) | ADNI | HC vs AD (117 / 113) HC vs MCI (117 / 110) MCInc vs MCIc (83 / 27) | GM volumes from T1w and uptake values from PET in 90 ROI's (FS by t-test). | PET (FDG and Florbetapir). | SRC | Sen = 0.956 / Spe = 0.940 Sen = 0.664 / Spe = 0.821 Sen = 0.741 / Spe = 0.815 |
| (Yang, Li et al. 2014) | ADNI | HC vs AD (150 / 70) HC vs MCI (150 / 98) | Voxel-wise values from GM map (ICA decomposition and FS by t-test). | MMSE, GDTOTAL, HMSCORE. | SVM | Sen = 0.992 / Spe = 0.962 Sen = 0.860 / Spe = 0.896 |
| (Ye, Pohl et al. 2011) | ADNI | MCInc vs MCIc (169 / 68) | GM RAVENS map (graph representation and FS by ISOMAP). | n.a. | SVM (linear Laplacian, semi-supervised) | Sen = 0.941 / Spe = 0.408 |
| (Ye, Zu et al. 2015) | ADNI | HC vs AD (52 / 51) HC vs MCI (52 / 99) | GM volumes and PET intensity values in 93 ROI's (FS by discriminative multi-task approach). | PET. | SVM (linear, multi-kernel) | Sen = 0.947 / Spe = 0.971 Sen = 0.877 / Spe = 0.715 |
| (Young, Modat et al. 2013) | ADNI | MCInc vs MCIc (96 / 47) | GM volumes and PET intensity values in 920 ROI's. | PET, APOE3. | Gaussian process | Sen = 0.787 / Spe = 0.656 |
| (Zhang, Wang et al. 2011) | ADNI | HC vs AD (52 / 51) HC vs MCI (52 / 99) | GM volumes and PET intensity values in 93 ROI's. | PET, CSF biomarkers. | SMV (linear, multi-kernel) | Sen = 0.930 / Spe = 0.933 Sen = 0.818 / Spe = 0.660 |
| (Zhang, Wang et al. 2015) | OASIS | HC vs MCI vs AD (97 / 57 / 24) | 3D-DWT decomposition features, ICV, atlas scaling factor, normalized brain volume (FS by PCA.) | Demographics, Education, SES, MMSE. | SMV (RBF) | Acc = 0.815 |
| (Zhang and Wang 2015) | OASIS | HC vs AD (98 / 28) | Voxel-wise displacement field values (direction and magnitude) of key T1w slices. (FS by PCA.) | PET, APOE3. | SVM (twin) | Sen = 0.906 / Spe = 0.934 |
| (Zhang, Stonnington et al. 2016) | ADNI | HC vs AD (228 / 194) MCInc vs MCIc (246 / 142) | Hippocampal surface tensor-based morphometry features and radial distance (FS by sparse coding) | n.a. | Adaboost | Sen = 0.830 / Spe = 0.780 Sen = 0.820 / Spe = 0.760 |
| (Zheng, Yao et al. 2015) | ADNI | HC vs AD (189 / 163) HC vs MCI (198 / 198) MCI vs AD (198 / 163) MCInc vs MCIc (94 / 104) | Cortical thickness in 78 ROI's (correlation matrix representation, FS by mRMR and SVM-RFE). | APOE3. | SVM (RBF) | Sen = 0.899 / Spe = 0.943 Sen = 0.878 / Spe = 0.858 Sen = 0.806 / Spe = 0.899 Sen = 0.789 / Spe = 0.799 |
| (Zheng, Shi et al. 2016) | ADNI | HC vs AD (52 / 51) | GM volumes and PET intensity values in 93 ROI's (high-level representation from multi-modality stacked deep polynomial network) | PET. | SVM (linear) | Sen = 0.973 / Spe = 0.983 |
| (Zhou, Goryawala et al. 2014) | ADNI | HC vs AD (127 / 59) HC vs MCIc (127 / 67) HC vs MCInc (127 / 56) | 41 regional and 10 morphometric volumes (FS by t-test). | MMSE. | SVM (RBF) | Sen = 0.840 / Spe = 0.961 Sen = 0.611 / Spe = 0.834 Sen = 0.552 / Spe = 0.823 |
| (Zhu, Suk et al. 2014) | ADNI | HC vs AD (52 / 51) HC vs MCI (52 / 99) | GM volumes and PET intensity values in 93 ROI's (FS by regularized least square regression). | n.a. | SVM | Sen = 0.886 / Spe = 0.978 Sen = 0.948 / Spe = 0.569 |
| (Zhu and Shi 2014) | ADNI | HC vs AD (52 / 51) | GM volumes in 93 ROI's (co-training semi-supervised learning approach). | n.a. | SVM (linear) | Sen = 0.869 / Spe = 0.904 |
| (Zhu, Shi et al. 2014) | ADNI | HC vs AD (52 / 51) | GM volumes and PET intensity values in 93 ROI's (FS by Hessian regularization semi-supervised approach). | n.a. | SVM (linear) | Sen = 0.952 / Spe = 0.907 |

| | | | | | | | |
|---|---|---|---|---|---|---|---|
| (Zhu, Suk et al. 2015) | ADNI | HC vs MCI vs AD (52 / 99 / 51) | | GM volumes and PET intensity values in 93 ROI's (CCA representation and FS by multi-task learning). | PET. | SMV | Acc = 0.729 |
| | | HC vs MCInc vs MCIc vs AD (52 / 56 / 43 / 51) | | | | | Acc = 0.619 |

## Table S4. Machine learning methods for classification of other types of dementia.

| Reference | Dataset (population) | Validation set size | Tasks | Imaging sequences | Imaging features (FS and representation) | Classifiers | Acc |
|---|---|---|---|---|---|---|---|
| (Chen, Tong et al. 2015) | Local (Stroke) | 350 / 240 | HC vs SVD | CT | Voxel intensities in ROI's from WML-based atlas. | MIL | 0.75 |
| (Koikkalainen, Rhodius-Meester et al. 2016) | Local (Dementia) | 118 / 223 / 92 / 47 / 24 | HC vs AD vs FTD vs DLB vs SVD | T1w, FLAIR. | Volumes of 142 ROI's, values of TBM and VBM in 140 ROI's, 20 manifold learning features, 8 ROI-based gradings and 1 vascular burden measure. | Multimodal statistical approach. | 0.706 |
| (Oppedal, Eftestøl et al. 2015) | Local (Dementia) | 36 / 57 / 16 | HC vs AD vs LBD | T1w, FLAIR. | Voxel-wise 2D-LBP and contrast features from WM and WML regions in T1w and FLAIR (FS by best first approach). | Random forest. | 0.87 |
| (Vemuri, Simon et al. 2011) | Local (Dementia) | 48 / 20 / 47 | AD vs FTD vs LBD | T1w. | GM volumes in 91 ROI's (FS by LDA). | k-means | 0.867 |
| (Wang, Redmond et al. 2016) | Local (Dementia) | 54 / 55 57 / 54 / 55 | AD vs FTD HC vs AD vs FTD | T1w. | 17 neurophysiological features and GM volumes of 8 ROI'ss (amygdala, hippocampus, medial temporal lobe, temporal pole, dorsolateral prefrontal cortex, ventromedial prefrontal cortex, striatum and insula (FS by best first approach). | Naïve Bayes | 0.6147 0.6747 |

**Table S5:** Machine learning studies on lesion segmentation; top, white matter hyperintensities; bottom, ischaemic stroke lesions. DC = DICE coefficient where a value close to 1 indicates perfect match of the test segmentation with the reference standard and 0 indicates no overlap.

| Reference | Dataset (population) | Validation set size | Target | Imaging sequences | Imaging features (FS and representation) | Classifiers | DC |
|---|---|---|---|---|---|---|---|
| (Fiot, Cohen et al. 2013) | Local (Ageing) | 125 | WMH | T1w, T2w, FLAIR, PD. | Neighbourhood voxel intensities and pyramidal features (Gaussian kernels) from each modality in ROI's. | SVM (RBF) | 0.69 |
| (Ithapu, Singh et al. 2014) | Local (AD and ageing) | 38 | WMH | T1w, FLAIR. | Voxel intensities and textons in ROI's. | Random forest | 0.67 |
| (Erus, Zacharaki et al. 2014) | Local (Diabetic and ageing) | 80 | WMH | FLAIR. | Voxel-wise intensity values of abnormality map in ROI's (FS by wavelet-based approach). | Iterative wavelet-based PCA model. | 0.59 |
| (Griffanti, Zamboni et al. 2016) | Local (TIA or minor stroke, no lacunar infarcts) | 82 | WMH | FLAIR. | Spatially weighted voxel-wise intensity values, patch average intensity. | kNN | 0.52 |
| (Vos, Biesbroek et al. 2013) | Local (Stroke) | 30 | Stroke | CT. | Location and volume of lesion, voxel intensities and likelihood of belonging to a lesion in lesion and mirrored region (FS by first best search approach). | Random forest | 0.74 |
| (Guo, Fridriksson et al. 2015) | Local (Stroke) | 60 | Stroke | T1w. | Voxel-wise 0-, 1-, 2-order statistical features from T1w, GM, WM, CSF and lesion probability map. | SVM (linear, ensemble) | 0.73 |
| (Maier, Schröder et al. 2015) | Local (spatial neglect) | 35 | Stroke | FLAIR. | Voxel intensity and location, weighted mean and histogram in voxel neighbourhood. | Random forest | 0.67 |

**Table S 6.** Machine learning for detection of specific (small) lesions; top, microbleeds (CMB); bottom, lacunes.

| Reference | Dataset (population) | Validation set size | Target | Imaging sequences | Imaging features (FS and representation) | Classifiers | Pre | Rec |
|---|---|---|---|---|---|---|---|---|
| (Ghafaryasl, van der Lijn et al. 2012) | Local (Ageing) | 81 | CMB | T2*, GRE. | Intensity, size and shape features from candidate ROI's in T2*. Intensity in GRE (FS by feed-forward approach). | Parzen, QDC | 0.352 | 0.99 |
| (Dou, Chen et al. 2016) | Local (Stroke and ageing) | 50 | CMB | SWI | 3D patches of SWI used as input. | 3D conv-NN | 0.443 | 0.93 |
| (Fazlollahi, Meriaudeau et al. 2015) | Local (Diabetic and ageing) | 66 | CMB | SWI | 3D Radon- and Hessian-based shape features from candidate ROI's. | Random forest (cascade) | 0.101 | 0.87 |
| (Uchiyama, Abe et al. 2014) | Local (lacunar infarcts) | 132 | Stroke | T1w, T2w | Location, intensity differences in T1w and T2w, multi-scale nodular and linear component (FS by PCA). | SVM | 0.154 | 0.97 |

Figure S1 QUADAS-2 charts of the studies included in the review

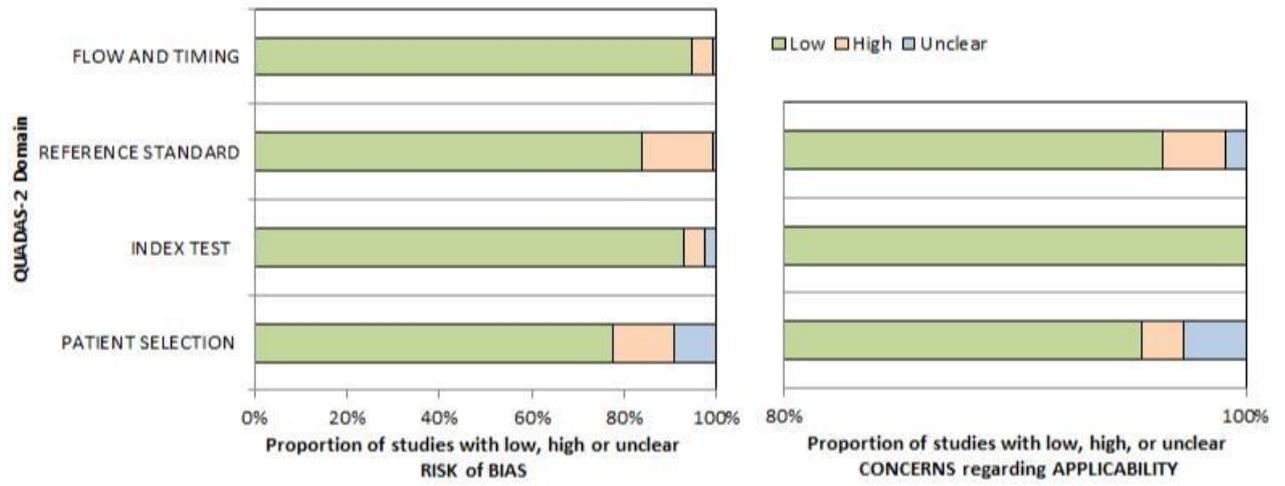

Figure S2 – Forest plot of accuracy of studies for differentiating different cognitive states ordered by data source, 1[st] page.

**Healthy controls v Alzheimers patients**

**Healthy controls v Mild Cognitive Impairment patients**

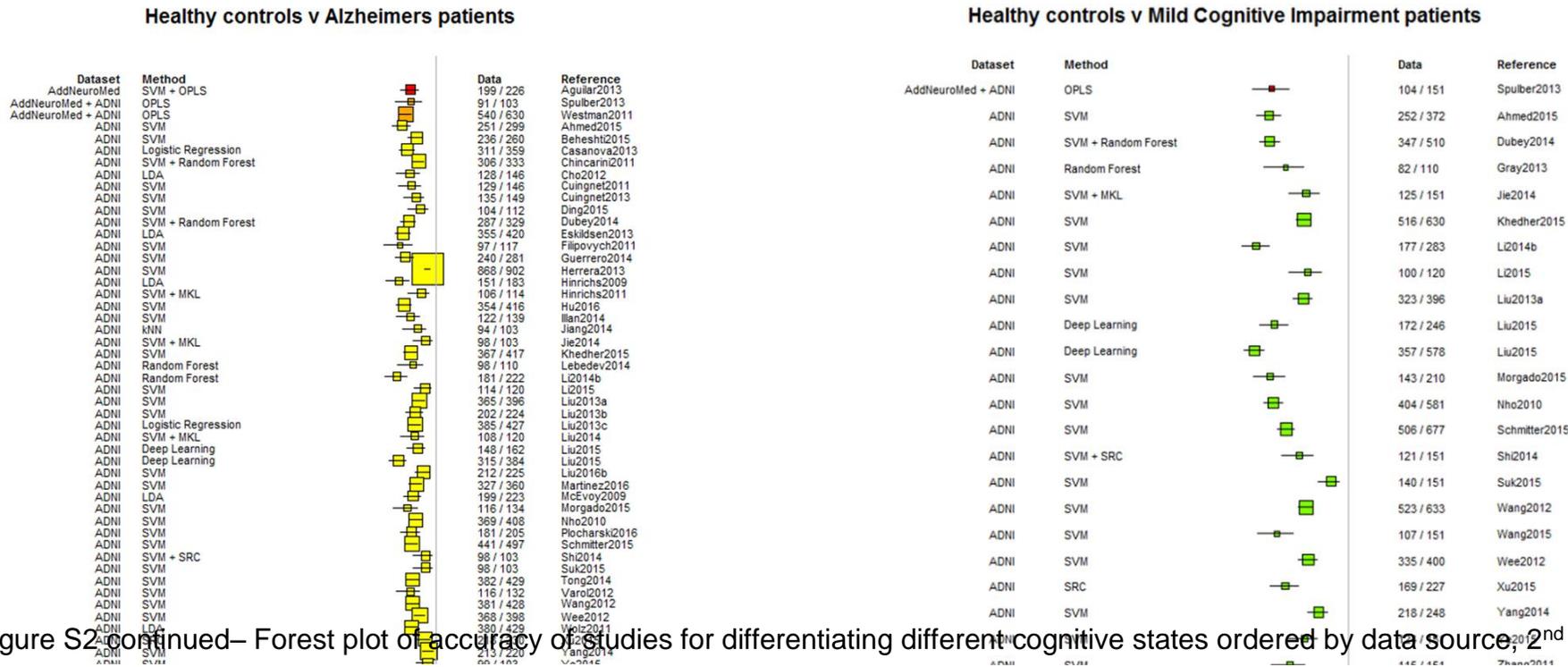

Figure S2 continued– Forest plot of accuracy of studies for differentiating different cognitive states ordered by data source, 2[nd] page.

**Non-converting v Converting Mild Cognitive Impairment patients**

**Mild Cognitive Impairment v Alzheimer's patients**

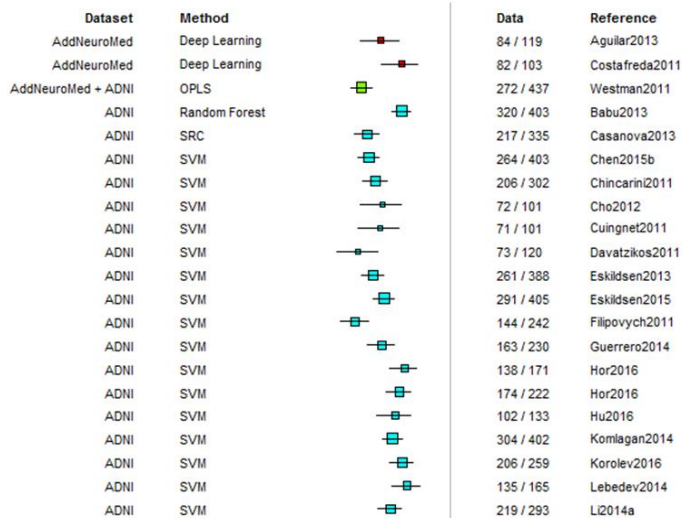

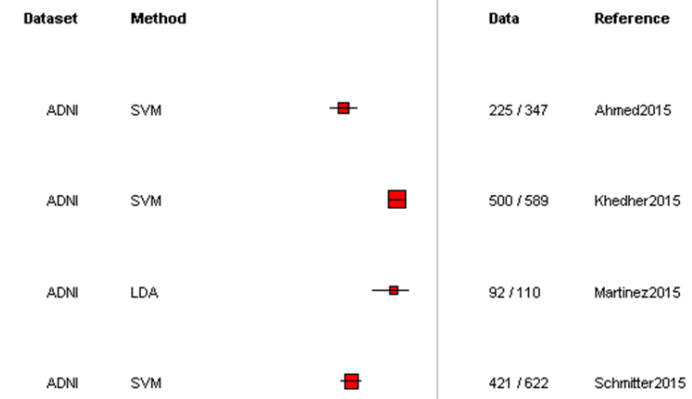

Figure S3 – Forest plot of accuracy of studies for differentiating different cognitive states ordered by machine learning method, 1<sup>st</sup> page.

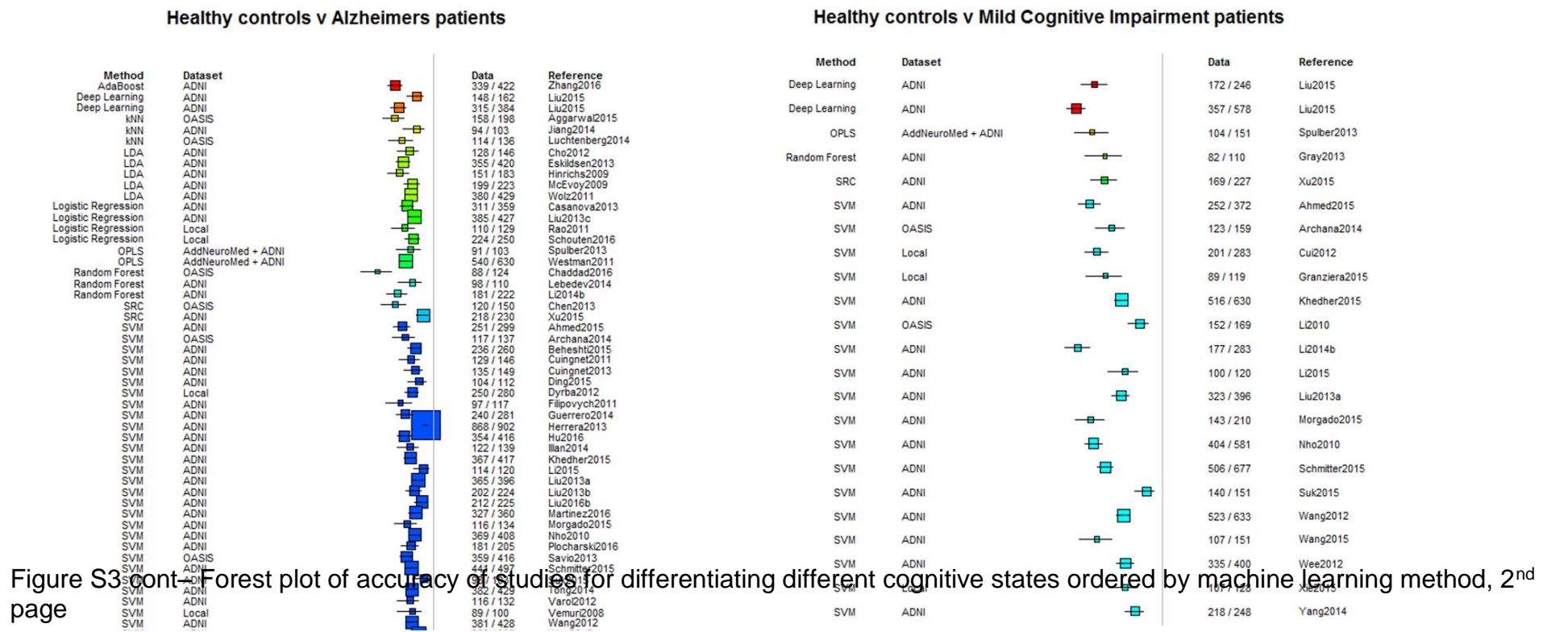

Figure S3 cont – Forest plot of accuracy of studies for differentiating different cognitive states ordered by machine learning method, 2<sup>nd</sup> page

Figure S4 – Forest plot of accuracy of studies for differentiating different cognitive states ordered by study size, 1st page.

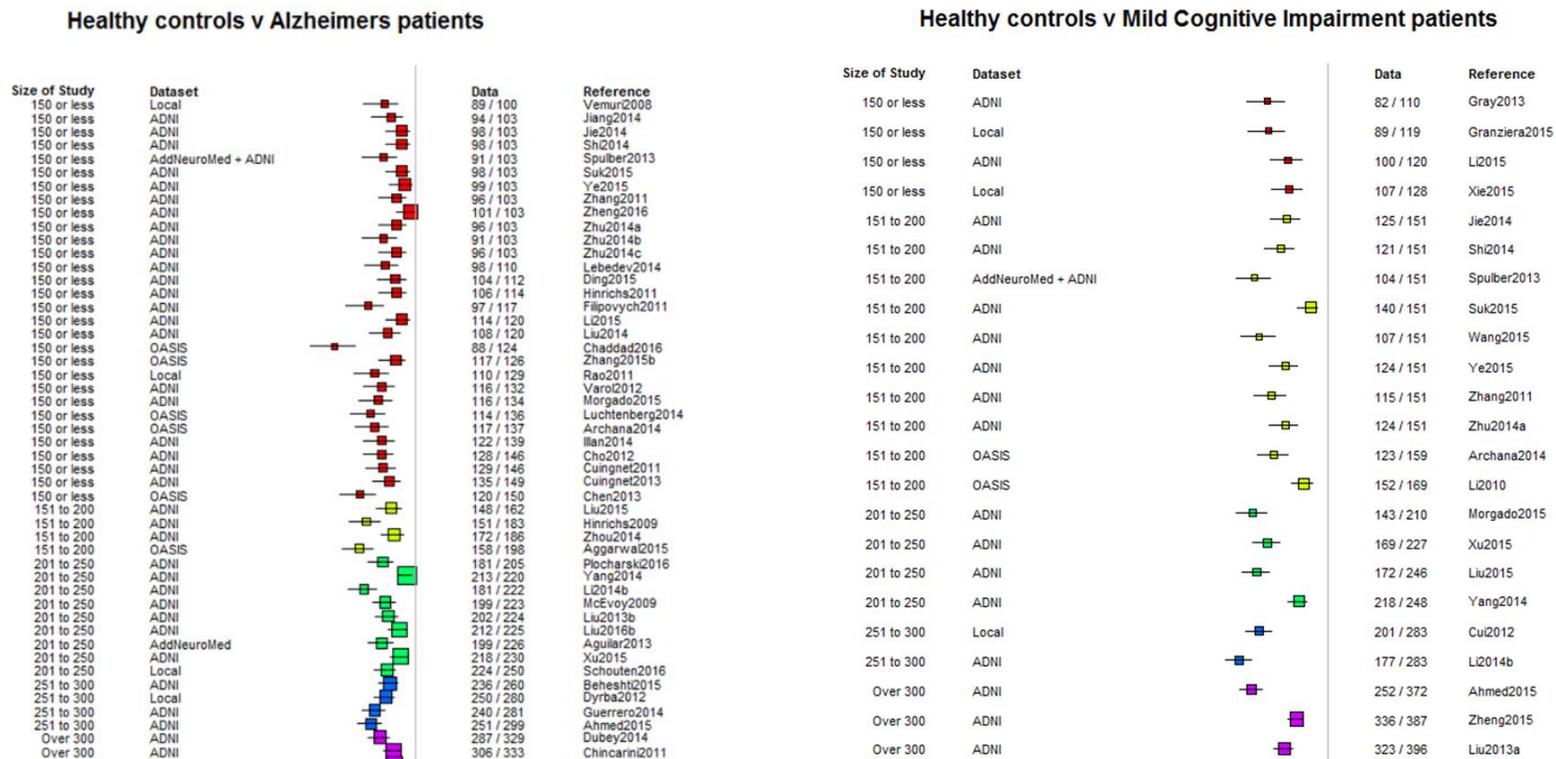

Figure S4 continued – Forest plot of accuracy of studies for differentiating different cognitive states ordered by study size, 2nd page.

**Non-converting v Converting Mild Cognitive Impairment patients**

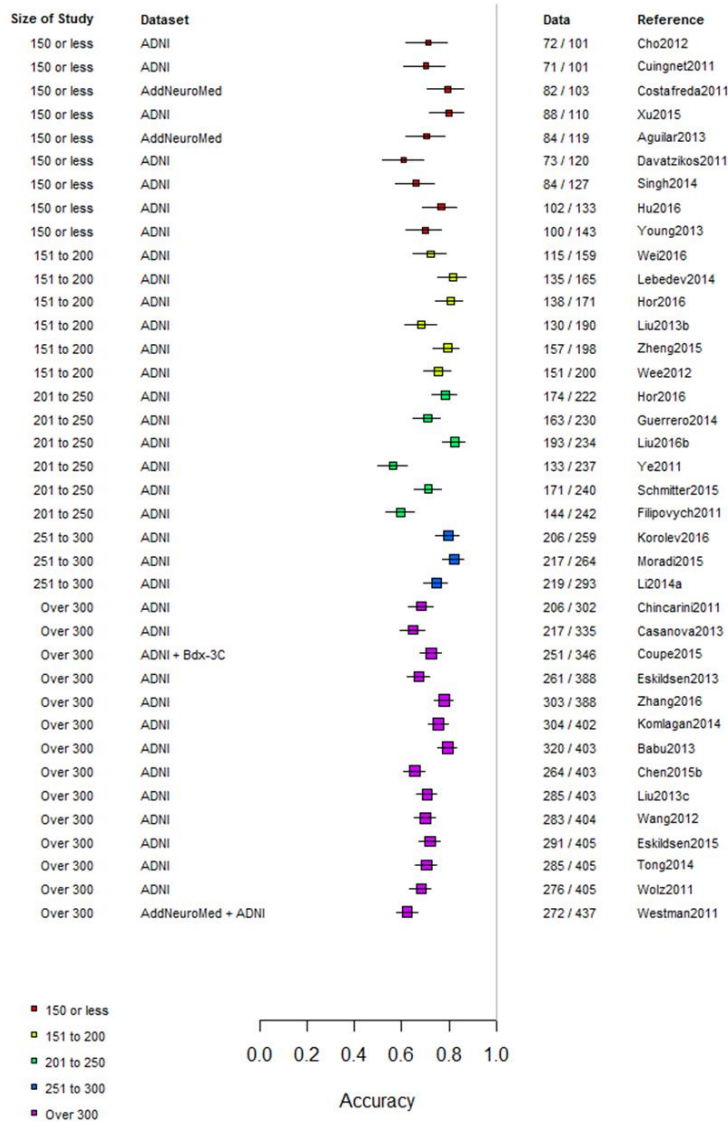

**Mild Cognitive Impairment v Alzheimer's patients**

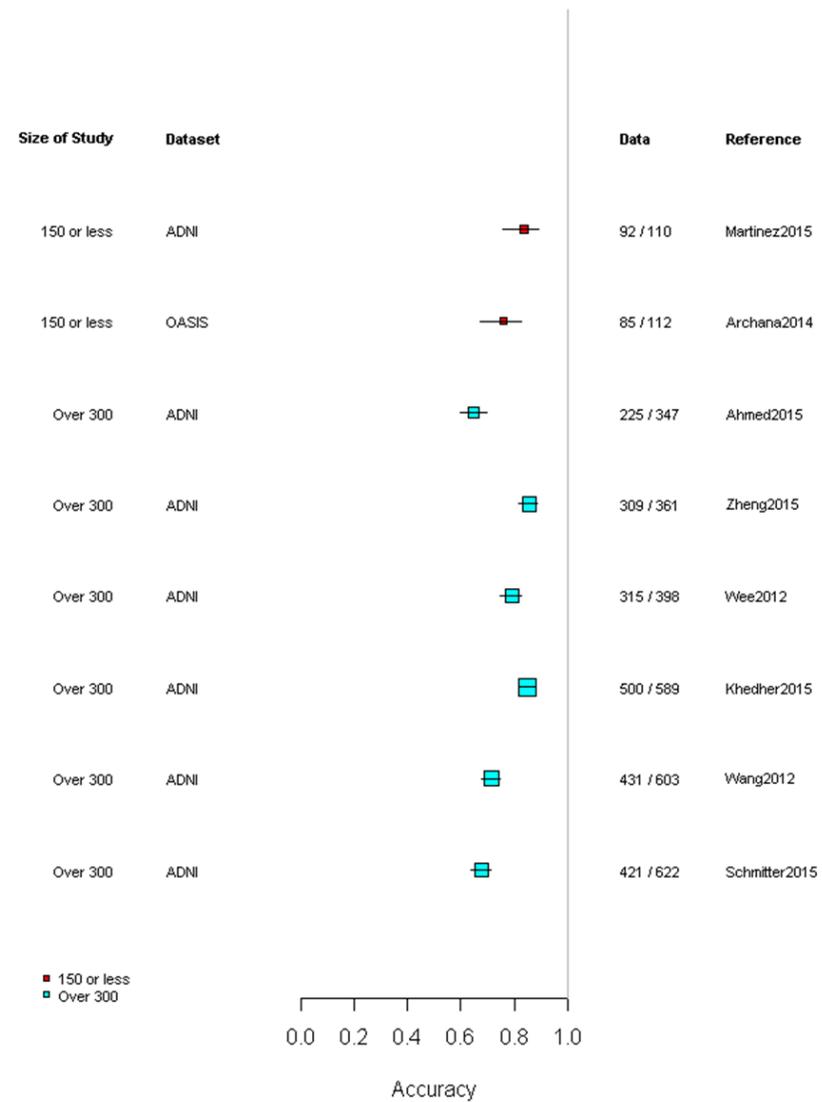

# References

Aggarwal, N., et al. (2015). "3d discrete wavelet transform for computer aided diagnosis of Alzheimer's disease using t1-weighted brain MRI." International Journal of Imaging Systems and Technology **25**: 179-190.

Aguilar, C., et al. (2013). "Different multivariate techniques for automated classification of MRI data in Alzheimer's disease and mild cognitive impairment." Psychiatry Research: Neuroimaging **212**: 89-98.

Ahmed, O. B., et al. (2015). "Alzheimer's disease diagnosis on structural MR images using circular harmonic functions descriptors on hippocampus and posterior cingulate cortex." Computerized Medical Imaging and Graphics **44**: 13-25.

Anagnostopoulos, C.-N., et al. (2013). Classification models for Alzheimer's disease detection. International Conference on Engineering Applications of Neural Networks**:** 193-202.

Archana, M. and S. Ramakrishnan (2014). Detection of Alzheimer disease in MR images using structure tensor. 2014 36th Annual International Conference of the IEEE Engineering in Medicine and Biology Society**:** 1043-1046.

Babu, G. S., et al. (2013). Meta-cognitive q-Gaussian RBF network for binary classification: Application to mild cognitive impairment (MCI). Neural Networks (IJCNN), The 2013 International Joint Conference on**:** 1-8.

Beheshti, I., et al. (2015). "Probability distribution function-based classification of structural MRI for the detection of Alzheimer's disease." Computers in biology and medicine **64**: 208-216.

Casanova, R., et al. (2013). "Alzheimer's disease risk assessment using large-scale machine learning methods." PLoS ONE **8**: e77949.

Chaddad, A., et al. (2016). Local discriminative characterization of MRI for Alzheimer's disease. 2016 IEEE 13th International Symposium on Biomedical Imaging (ISBI), Institute of Electrical and Electronics Engineers (IEEE).

Chen, L., et al. (2015). Identification of Cerebral Small Vessel Disease Using Multiple Instance Learning. International Conference on Medical Image Computing and Computer-Assisted Intervention**:** 523-530.

Chen, X., et al. (2015). Group Sparse Representation for Prediction of MCI Conversion to AD. International Conference on Intelligent Computing**:** 510-519.

Chen, Y. and T. D. Pham (2013). "Development of a brain MRI-based hidden Markov model for dementia recognition." Biomedical engineering online **12**: 1.

Chincarini, A., et al. (2011). "Local MRI analysis approach in the diagnosis of early and prodromal Alzheimer's disease." Neuroimage **58**: 469-480.

Cho, Y., et al. (2012). "Individual subject classification for Alzheimer's disease based on incremental learning using a spatial frequency representation of cortical thickness data." Neuroimage **59**: 2217-2230.

Costafreda, S. G., et al. (2011). "Automated hippocampal shape analysis predicts the onset of dementia in mild cognitive impairment." NeuroImage **56**: 212-219.

Coupé, P., et al. (2015). "Detection of Alzheimer's disease signature in MR images seven years before conversion to dementia: Toward an early individual prognosis." Human Brain Mapping **36**: 4758-4770.

Cui, Y., et al. (2012). "Automated detection of amnestic mild cognitive impairment in community-dwelling elderly adults: a combined spatial atrophy and white matter alteration approach." Neuroimage **59**: 1209-1217.

Cuingnet, R., et al. (2011). "Automatic classification of patients with Alzheimer's disease from structural MRI: a comparison of ten methods using the ADNI database." neuroimage **56**: 766-781.

Cuingnet, R., et al. (2013). "Spatial and Anatomical Regularization of SVM: A General Framework for Neuroimaging Data." IEEE Transactions on Pattern Analysis and Machine Intelligence **35**: 682-696.

Davatzikos, C., et al. (2011). "Prediction of MCI to AD conversion, via MRI, CSF biomarkers, and pattern classification." Neurobiology of aging **32**: 2322--e2319.

Ding, Y., et al. (2015). Classification of Alzheimer's disease based on the combination of morphometric feature and texture feature. Bioinformatics and Biomedicine (BIBM), 2015 IEEE International Conference on**:** 409-412.

Dou, Q., et al. (2016). "Automatic Detection of Cerebral Microbleeds From MR Images via 3D Convolutional Neural Networks." <u>IEEE Trans Med Imaging</u> **35**: 1182-1195.

Dubey, R., et al. (2014). "Analysis of sampling techniques for imbalanced data: An n= 648 ADNI study." <u>NeuroImage</u> **87**: 220-241.

Dyrba, M., et al. (2012). Combining DTI and MRI for the automated detection of Alzheimer's disease using a large European multicenter dataset. <u>International Workshop on Multimodal Brain Image Analysis</u>**:** 18-28.

Erus, G., et al. (2014). "Individualized statistical learning from medical image databases: Application to identification of brain lesions." <u>Medical image analysis</u> **18**: 542-554.

Eskildsen, S. F., et al. (2015). "Structural imaging biomarkers of Alzheimer's disease: predicting disease progression." <u>Neurobiology of aging</u> **36**: S23--S31.

Eskildsen, S. F., et al. (2013). "Prediction of Alzheimer's disease in subjects with mild cognitive impairment from the ADNI cohort using patterns of cortical thinning." <u>Neuroimage</u> **65**: 511-521.

Fazlollahi, A., et al. (2015). "Computer-aided detection of cerebral microbleeds in susceptibility-weighted imaging." <u>Computerized Medical Imaging and Graphics</u> **46**: 269-276.

Filipovych, R., et al. (2011). "Semi-supervised pattern classification of medical images: application to mild cognitive impairment (MCI)." <u>NeuroImage</u> **55**: 1109-1119.

Fiot, J.-B., et al. (2013). "Efficient brain lesion segmentation using multi-modality tissue-based feature selection and support vector machines." <u>International Journal for Numerical Methods in Biomedical Engineering</u> **29**: 905-915.

Ghafaryasl, B., et al. (2012). A computer aided detection system for cerebral microbleeds in brain MRI. <u>2012 9th IEEE International Symposium on Biomedical Imaging (ISBI)</u>**:** 138-141.

Granziera, C., et al. (2015). "A multi-contrast MRI study of microstructural brain damage in patients with mild cognitive impairment." <u>NeuroImage: Clinical</u> **8**: 631-639.

Gray, K. R., et al. (2013). "Random forest-based similarity measures for multi-modal classification of Alzheimer's disease." <u>NeuroImage</u> **65**: 167-175.

Griffanti, L., et al. (2016). "BIANCA (Brain Intensity AbNormality Classification Algorithm): A new tool for automated segmentation of white matter hyperintensities." <u>NeuroImage</u> **141**: 191-205.

Guerrero, R., et al. (2014). "Manifold population modeling as a neuro-imaging biomarker: Application to ADNI and ADNI-GO." <u>NeuroImage</u> **94**: 275-286.

Guo, D., et al. (2015). "Automated lesion detection on MRI scans using combined unsupervised and supervised methods." <u>BMC medical imaging</u> **15**: 1.

Herrera, L. J., et al. (2013). Classification of MRI Images for Alzheimer's Disease Detection. <u>2013 International Conference on Social Computing</u>, Institute of Electrical and Electronics Engineers (IEEE).

Hinrichs, C., et al. (2009). "Spatially augmented LPboosting for AD classification with evaluations on the ADNI dataset." <u>NeuroImage</u> **48**: 138-149.

Hinrichs, C., et al. (2011). "Predictive markers for AD in a multi-modality framework: An analysis of MCI progression in the ADNI population." <u>NeuroImage</u> **55**: 574-589.

Hor, S. and M. Moradi (2016). "Learning in data-limited multimodal scenarios: Scandent decision forests and tree-based features." <u>Medical Image Analysis</u> **34**: 30-41.

Hu, K., et al. (2016). "Multi-scale features extraction from baseline structure MRI for MCI patient classification and AD early diagnosis." <u>Neurocomputing</u> **175**: 132-145.

Illan, I. A., et al. (2014). "Spatial component analysis of MRI data for Alzheimer's disease diagnosis: a Bayesian network approach." <u>Front. Comput. Neurosci.</u> **8**.

Ithapu, V., et al. (2014). "Extracting and summarizing white matter hyperintensities using supervised segmentation methods in Alzheimer's disease risk and aging studies." Human Brain Mapping: n/a--n/a.

Jiang, Q. and J. Shi (2014). Sparse kernel entropy component analysis for dimensionality reduction of neuroimaging data. 2014 36th Annual International Conference of the IEEE Engineering in Medicine and Biology Society, Institute of Electrical and Electronics Engineers (IEEE).

Jie, B., et al. (2014). "Manifold regularized multitask feature learning for multimodality disease classification." Human Brain Mapping 36: 489-507.

Khedher, L., et al. (2015). "Early diagnosis of Alzheimer's disease based on partial least squares, principal component analysis and support vector machine using segmented MRI images." Neurocomputing 151: 139-150.

Koikkalainen, J., et al. (2016). "Differential diagnosis of neurodegenerative diseases using structural MRI data." NeuroImage: Clinical 11: 435-449.

Komlagan, M., et al. (2014). Anatomically Constrained Weak Classifier Fusion for Early Detection of Alzheimer's Disease. International Workshop on Machine Learning in Medical Imaging: 141-148.

Korolev, I. O., et al. (2016). "Predicting Progression from Mild Cognitive Impairment to Alzheimer's Dementia Using Clinical, MRI, and Plasma Biomarkers via Probabilistic Pattern Classification." PLoS ONE 11: e0138866.

Krashenyi, I., et al. (2016). "Fuzzy Computer-Aided Alzheimer's Disease Diagnosis Based on MRI Data." Current Alzheimer Research 13: 545-556.

Lebedev, A. V., et al. (2014). "Random Forest ensembles for detection and prediction of Alzheimer's disease with a good between-cohort robustness." NeuroImage: Clinical 6: 115-125.

Li, H., et al. (2014). "Hierarchical Interactions Model for Predicting Mild Cognitive Impairment (MCI) to Alzheimer's Disease (AD) Conversion." PLoS ONE 9: e82450.

Li, L., et al. (2010). Detection of Mild Cognitive Impairment Using Image Differences and Clinical Features. 2010 IEEE International Conference on BioInformatics and BioEngineering, Institute of Electrical and Electronics Engineers (IEEE).

Li, M., et al. (2014). "An Efficient Approach for Differentiating Alzheimer's Disease from Normal Elderly Based on Multicenter MRI Using Gray-Level Invariant Features." PLoS ONE 9: e105563.

Li, Y., et al. (2015). Classification of Alzheimer's Disease Based on Multiple Anatomical Structures' Asymmetric Magnetic Resonance Imaging Feature Selection. Neural Information Processing, Springer Science Business Media: 280-289.

Liu, F., et al. (2013). High-order graph matching based feature selection for Alzheimer's disease identification. International Conference on Medical Image Computing and Computer-Assisted Intervention: 311-318.

Liu, F., et al. (2014). "Multiple Kernel Learning in the Primal for Multimodal Alzheimer's Disease Classification." IEEE Journal of Biomedical and Health Informatics 18: 984-990.

Liu, M., et al. (2016). "Inherent Structure-Based Multiview Learning With Multitemplate Feature Representation for Alzheimer's Disease Diagnosis." IEEE Transactions on Biomedical Engineering 63: 1473-1482.

Liu, M., et al. (2013). "Identifying Informative Imaging Biomarkers via Tree Structured Sparse Learning for AD Diagnosis." Neuroinform 12: 381-394.

Liu, S., et al. (2016). "Cross-View Neuroimage Pattern Analysis in Alzheimer's Disease Staging." Frontiers in Aging Neuroscience 8.

Liu, S., et al. (2015). "Multimodal Neuroimaging Feature Learning for Multiclass Diagnosis of Alzheimer's Disease." IEEE Transactions on Biomedical Engineering 62: 1132-1140.

Liu, X., et al. (2013). "Locally linear embedding (LLE) for MRI based Alzheimer's disease classification." NeuroImage 83: 148-157.

Luchtenberg, A., et al. (2014). Early detection of Alzheimer's disease using histograms in a dissimilarity-based classification framework. SPIE Medical Imaging: 903502.

Maier, O., et al. (2015). "Classifiers for Ischemic Stroke Lesion Segmentation: A Comparison Study." PLoS ONE 10: e0145118.

Martinez-Murcia, F., et al. (2016). "A Spherical Brain Mapping of MR Images for the Detection of Alzheimer's Disease." CAR **13**: 575-588.

Martinez-Torteya, A., et al. (2015). "Improved Diagnostic Multimodal Biomarkers for Alzheimer's Disease and Mild Cognitive Impairment." BioMed Research International **2015**: 1-11.

McEvoy, L. K., et al. (2009). "Alzheimer Disease: Quantitative Structural Neuroimaging for Detection and Prediction of Clinical and Structural Changes in Mild Cognitive Impairment." Radiology **251**: 195-205.

Moradi, E., et al. (2015). "Machine learning framework for early MRI-based Alzheimer's conversion prediction in MCI subjects." NeuroImage **104**: 398-412.

Morgado, P. M. and M. Silveira (2015). "Minimal neighborhood redundancy maximal relevance: Application to the diagnosis of Alzheimer's disease." Neurocomputing **155**: 295-308.

Nho, K., et al. (2010). Automatic prediction of conversion from mild cognitive impairment to probable Alzheimer's disease using structural magnetic resonance imaging. AMIA Annual Symposium Proceedings. **2010:** 542.

Oppedal, K., et al. (2015). "Classifying Dementia Using Local Binary Patterns from Different Regions in Magnetic Resonance Images." International Journal of Biomedical Imaging **2015**: 1-14.

Plocharski, M. and L. R. Østergaard (2016). "Extraction of sulcal medial surface and classification of Alzheimer's disease using sulcal features." Computer Methods and Programs in Biomedicine **133**: 35-44.

Rao, A., et al. (2011). Classification of Alzheimer's Disease from structural MRI using sparse logistic regression with optional spatial regularization. 2011 Annual International Conference of the IEEE Engineering in Medicine and Biology Society, Institute of Electrical and Electronics Engineers (IEEE).

Rueda, A., et al. (2014). "Extracting Salient Brain Patterns for Imaging-Based Classification of Neurodegenerative Diseases." IEEE Trans Med Imaging **33**: 1262-1274.

Savio, A. and M. GrañA (2013). "Deformation based feature selection for computer aided diagnosis of Alzheimer's disease." Expert Systems with Applications **40**: 1619-1628.

Schmitter, D., et al. (2015). "An evaluation of volume-based morphometry for prediction of mild cognitive impairment and Alzheimer's disease." NeuroImage: Clinical **7**: 7-17.

Schouten, T. M., et al. (2016). "Combining anatomical, diffusion, and resting state functional magnetic resonance imaging for individual classification of mild and moderate Alzheimer's disease." NeuroImage: Clinical **11**: 46-51.

Shi, Y., et al. (2014). Joint Coupled-Feature Representation and Coupled Boosting for AD Diagnosis. 2014 IEEE Conference on Computer Vision and Pattern Recognition, Institute of Electrical and Electronics Engineers (IEEE).

Singh, N., et al. (2014). "Quantifying anatomical shape variations in neurological disorders." Medical Image Analysis **18**: 616-633.

Spulber, G., et al. (2013). "An MRI-based index to measure the severity of Alzheimer's disease-like structural pattern in subjects with mild cognitive impairment." J Intern Med **273**: 396-409.

Suk, H.-I., et al. (2015). "Deep sparse multi-task learning for feature selection in Alzheimer's disease diagnosis." Brain Structure and Function **221**: 2569-2587.

Tong, T., et al. (2014). "Multiple instance learning for classification of dementia in brain MRI." Medical Image Analysis **18**: 808-818.

Uchiyama, Y., et al. (2014). "Eigenspace Template Matching for Detection of Lacunar Infarcts on MR Images." Journal of Digital Imaging **28**: 116-122.

Varol, E., et al. (2012). Feature ranking based nested support vector machine ensemble for medical image classification. 2012 9th IEEE International Symposium on Biomedical Imaging (ISBI), Institute of Electrical and Electronics Engineers (IEEE).

Vemuri, P., et al. (2008). "Alzheimer's disease diagnosis in individual subjects using structural MR images: Validation studies." NeuroImage **39**: 1186-1197.

Vemuri, P., et al. (2011). "Antemortem differential diagnosis of dementia pathology using structural MRI: Differential-STAND." NeuroImage **55**: 522-531.

Vos, P. C., et al. (2013). Automatic detection and segmentation of ischemic lesions in computed tomography images of stroke patients. Medical Imaging 2013: Computer-Aided Diagnosis. C. L. Novak and S. Aylward, SPIE-Intl Soc Optical Eng.

Wachinger, C. and M. Reuter (2016). "Domain adaptation for Alzheimer's disease diagnostics." NeuroImage **139**: 470-479.

Wang, B., et al. (2015). A hierarchical model for identifying mild cognitive impairment. 2015 11th International Conference on Natural Computation (ICNC), Institute of Electrical and Electronics Engineers (IEEE).

Wang, J., et al. (2016). "A Comparison of Magnetic Resonance Imaging and Neuropsychological Examination in the Diagnostic Distinction of Alzheimer's Disease and Behavioral Variant Frontotemporal Dementia." Frontiers in Aging Neuroscience **8**.

Wang, Y., et al. (2012). Groupwise segmentation improves neuroimaging classification accuracy. International Workshop on Multimodal Brain Image Analysis**:** 185-193.

Wee, C.-Y., et al. (2012). "Prediction of Alzheimer's disease and mild cognitive impairment using cortical morphological patterns." Human Brain Mapping **34**: 3411-3425.

Wei, R., et al. (2016). "Prediction of Conversion from Mild Cognitive Impairment to Alzheimer's Disease Using MRI and Structural Network Features." Frontiers in Aging Neuroscience **8**.

Westman, E., et al. (2011). "AddNeuroMed and ADNI: Similar patterns of Alzheimer's atrophy and automated MRI classification accuracy in Europe and North America." NeuroImage **58**: 818-828.

Wolz, R., et al. (2011). "Multi-Method Analysis of MRI Images in Early Diagnostics of Alzheimer's Disease." PLoS ONE **6**: e25446.

Xie, Y., et al. (2015). "Identification of Amnestic Mild Cognitive Impairment Using Multi-Modal Brain Features: A Combined Structural MRI and Diffusion Tensor Imaging Study." Journal of Alzheimer?s Disease **47**: 509-522.

Xu, L., et al. (2015). "Multi-modality sparse representation-based classification for Alzheimer's disease and mild cognitive impairment." Computer Methods and Programs in Biomedicine **122**: 182-190.

Yang, W., et al. (2014). ICA image feature extraction for improving diagnosis of Alzheimer's disease and mild cognitive impairment. 2014 10th International Conference on Natural Computation (ICNC), Institute of Electrical and Electronics Engineers (IEEE).

Ye, D. H., et al. (2011). Semi-supervised Pattern Classification: Application to Structural MRI of Alzheimer's Disease. 2011 International Workshop on Pattern Recognition in NeuroImaging, Institute of Electrical and Electronics Engineers (IEEE).

Ye, T., et al. (2015). Discriminative Multi-task Feature Selection for Multi-modality Based AD/MCI Classification. 2015 International Workshop on Pattern Recognition in NeuroImaging, Institute of Electrical and Electronics Engineers (IEEE).

Young, J., et al. (2013). "Accurate multimodal probabilistic prediction of conversion to Alzheimer's disease in patients with mild cognitive impairment." NeuroImage: Clinical **2**: 735-745.

Zhang, D., et al. (2011). "Multimodal classification of Alzheimer's disease and mild cognitive impairment." NeuroImage **55**: 856-867.

Zhang, J., et al. (2016). Applying sparse coding to surface multivariate tensor-based morphometry to predict future cognitive decline. 2016 IEEE 13th International Symposium on Biomedical Imaging (ISBI), Institute of Electrical and Electronics Engineers (IEEE).

Zhang, Y. and S. Wang (2015). "Detection of Alzheimer's disease by displacement field and machine learning." PeerJ **3**: e1251.

Zhang, Y., et al. (2015). "Detection of Alzheimer's disease and mild cognitive impairment based on structural volumetric MR images using 3D-DWT and WTA-KSVM trained by PSOTVAC." Biomedical Signal Processing and Control **21**: 58-73.

Zheng, W., et al. (2015). "Novel Cortical Thickness Pattern for Accurate Detection of Alzheimer's Disease." Journal of Alzheimer?s Disease **48**: 995-1008.

Zheng, X., et al. (2016). Multi-modality stacked deep polynomial network based feature learning for Alzheimer's disease diagnosis. 2016 IEEE 13th International Symposium on Biomedical Imaging (ISBI), Institute of Electrical and Electronics Engineers (IEEE).

Zhou, Q., et al. (2014). "An Optimal Decisional Space for the Classification of Alzheimer's Disease and Mild Cognitive Impairment." IEEE Transactions on Biomedical Engineering **61**: 2245-2253.

Zhu, J. and J. Shi (2014). Hessian regularization based semi-supervised dimensionality reduction for neuroimaging data of Alzheimer's disease. 2014 IEEE 11th International Symposium on Biomedical Imaging (ISBI), Institute of Electrical and Electronics Engineers (IEEE).

Zhu, J., et al. (2014). Co-training based semi-supervised classification of Alzheimer's disease. 2014 19th International Conference on Digital Signal Processing, Institute of Electrical and Electronics Engineers (IEEE).

Zhu, X., et al. (2015). "Canonical feature selection for joint regression and multi-class identification in Alzheimer's disease diagnosis." Brain Imaging and Behavior **10**: 818-828.

Zhu, X., et al. (2014). A Novel Multi-relation Regularization Method for Regression and Classification in AD Diagnosis. Medical Image Computing and Computer-Assisted Intervention MICCAI 2014, Springer Science Business Media**:** 401-408.